\documentclass[a4paper,11pt]{article}
\pdfoutput=1
\usepackage{jcappub,graphicx,slashed,MnSymbol}
\newcommand{\be}{\begin{equation}}
\newcommand{\ee}{\end{equation}}
\newcommand{\bea}{\begin{eqnarray}}
\newcommand{\eea}{\end{eqnarray}}
\newcommand{\DEWSB}{dynamical electroweak symmetry breaking}
\newcommand{\xx}{\check}
\begin{document}
\title{Strong interactions in air showers}
\author{Dennis D.~Dietrich}
\affiliation{Arnold Sommerfeld Center, Ludwig-Maximilians-Universit\"at, M\"unchen, Germany}
\affiliation{Institut f\"ur Theoretische Physik, Goethe-Universit\"at,  
Frankfurt am Main, Germany}
\date{\today}
\abstract{
We study the role new gauge interactions in extensions of the standard model play in air showers initiated by ultrahigh-energy cosmic rays.
Hadron-hadron events remain dominated by quantum chromodynamics, while projectiles and/or targets from beyond the standard model permit us to see qualitative differences arising due to the new interactions.
}
\maketitle


\section{Introduction}
The present study is concerned with the quest after the origin of mass in our universe and the phenomenology of ultrahigh-energy cosmic rays: On one hand, the huge energies available in cosmic rays might give evidence for \DEWSB. On the other hand, some observed cosmic-ray characteristics might be natural in the framework of \DEWSB. 

~

The standard model parametrises the masses of all known elementary particles with the help of the elementary scalar Higgs mechanism. Recently, experiments at the CERN LHC observed a scalar resonance that, so far, cannot be distinguished from the standard-model Higgs; thus confirming the standard model as effective description of nature.
All the other scalar degrees of freedom that have been encountered until now, however, turned out to be composites.
This is, for example, true for the scalar of Ginzburg and Landau \cite{Ginzburg:1950sr}, which describes the generation of the photon mass (the inverse London screening length \cite{London:1935}) in superconductivity in a gauge invariant manner and thus the Meissner-Ochsenfeld \cite{Meissner:1933} effect. It is the Abelian prototype for the standard-model Higgs, but was identified by Bardeen, Cooper, and Schrieffer (BCS) \cite{Cooper:1956} to be a composite of two electrons, the Cooper pair. For more guidance consider the standard model without the Higgs sector. There the chiral symmetry breaking in quantum chromodynamics (QCD) would break the electroweak symmetry dynamically. (In order to avoid the subtleties arising from the alignment of the vacuum, for this gedanken experiment, we constrain ourselves to one generation of fermions, especially to only two quarks.) With the scale of quantum chromodynamics kept fixed this would result in massive weak gauge bosons, which are roughly 1000 times to light. The pions of quantum chromodynamics would become the longitudinal degrees of freedom of the weak gauge bosons, and would, thus, not appear in the spectrum. This is suggestive of employing an additional sector of electroweakly charged fermions that interact under a new force that grows strong at low energy scales, leads to chiral symmetry breaking among the fermions, and, simultaneously, breaks the electroweak symmetry. This approach is known as technicolour \cite{Susskind:1978ms}, which is to the elementary Higgs mechanism what the BCS theory is to the parametrisation by Ginzburg and Landau. 

As will be explained in detail in Sect.~\ref{sec:dewsb}, the difference between theories of dynamical symmetry breaking and those with some kind of elementary Higgs mechanism that will be crucial vis-\`a-vis cosmic-ray physics, is the presence of additional gauge interactions. In that section, we are also going to discuss several different implementations of the dynamical-electroweak-symmetry-breaking paradigm in the context of cosmic-ray physics.

~

The dynamic electroweak symmetry breaking mechanism can manifest itself directly only at LHC energies or above. Hence, if we would like to see its traces we have to look at features at the high-energy end of the cosmic-ray spectrum, i.e., at cosmic-ray energies of at least 10$^{18}$eV. 
The most prominent feature below this energy range is the ``knee", an abrupt steepening of the spectrum between 10$^{15}$ and 10$^{16}$eV. (Here and in the following, see, for example, Fig.~27.8 in \cite{Beringer:1900zz}.) Under the assumption that all the cosmic rays up to 10$^{18}$eV have their origin inside the Milky Way, the knee could be the manifestation of the fact that all possible acceleration mechanisms within our galaxy have reached their maximal attainable energy.

From 10$^{18}$eV onward cosmic rays are expected to be of extragalactic origin. In an interval of one order of magnitude around 10$^{19}$eV the spectrum is flatter than after the knee. The place where this plateau begins is known as the ``ankle.'' In case one merely extrapolates cosmic-ray--air cross sections straightforwardly to higher energies, the ankle structure must be explained by a changing composition of the cosmic rays. Under this assumption, the direction of change (to lighter or to heavier particles), however, inferred by the HiRes \cite{Abbasi:2005} and Auger \cite{Unger:2007} experiments, respectively, from the depth at which the shower maximum occurs in the atmosphere, do not agree. Conclusions about the composition change considerably once the assumption of the validity of a straightforward extrapolation of the scattering-cross sections is dropped \cite{Beringer:1900zz}. Another explanation attempt for the ankle feature is a higher energy, e.g., cosmological flux that overtakes one of lower energy, e.g., a galactic flux \cite{Beringer:1900zz}, which, however, also would not overcome the composition issue.

At energies beyond $5\times10^{19}$eV the spectrum should decay rapidly if the cosmic rays were of extragalactic origin because, according to Greisen, Zatsepin, and Kuzmin (GZK) \cite{Greisen:1966} at these energies nuclei should lose a considerable fraction of their energy by pion production after scattering with photons from the cosmic microwave background (CMB) when propagating over extragalactic distances. (The analogous production of electron-positron pairs sets in at lower energies around 10$^{17}$eV, but is much less efficient in dispersing the cosmic ray energy than the heavier pions.) 
Trans-GZK events have been observed \cite{Bird:1994,Takeda:2003,Abbasi:2008}. The hypothesis of an isotropic distribution of the highest-energy cosmic rays is consistent with observations, and a correlation with the position of known active galactic nuclei, which would give an indication for their origin, is not confirmed \cite{Tsunesada:2011}. The AGASA experiment does not confirm the GZK suppression \cite{Takeda:2003}, whereas other experiments see some suppression \cite{Abbasi:2008,Tsunesada:2011,Abreu:2011}. 
Even if a steepening decay of the flux spectrum above the predicted GZK energy would be confirmed, a clear connection with the GZK mechanism could only be established by observing the particles radiated in the process that leads to the GZK cutoff, or their decay products. (As an example, half of the energy lost by protons in this way ends up in neutrinos \cite{Beresinsky:1969qj}.) Otherwise the suppression of the observed flux could be simply due to the local accelerating mechanism having a maximal attainable energy.
If the GZK mechanism turns out to be potent for ordinary hadrons,
the explanation of trans-GZK events requires either so far unknown {\sl galactic} sources or particles that are not affected by the GZK cutoff. These would have to be sufficiently weakly interacting particles from cosmological sources, which still have a cross section that allows them to be accelerated to the required energies and to interact with the atmosphere in order to be detected. 
Inversely, the confirmed absence of trans-GZK events together with the confirmed efficiency of the GZK mechanism (i.e., there are acceleration mechanisms that can bring particles to trans-GZK energies, but hadrons are decelerated efficiently by GZK processes), would put limits on the stability of such particles.

The rest of the paper is organised as follows. 
In section \ref{sec:dewsb} we recall those details about \DEWSB~we need in order to proceed with our analysis. Thereafter, in section \ref{sec:main} we describe in detail the effect of \DEWSB~expected for the interaction of hadronic cosmic-ray particles with air. In subsection \ref{sec:bsm}, we also allow for a component of the cosmic-ray flux that is not made from standard model particles, but from particles that are associated with \DEWSB. In the second part of that subsection, we extend our considerations also to cases, where the air shower is initiated by such a particle which is struck by a cosmic-ray particle. Finally, in Sect.~\ref{sec:sum} we summarise our findings.


\section{Dynamical electroweak symmetry breaking\label{sec:dewsb}}

Let us start with technicolour.
Fortunately, available data allows us to narrow down the set of viable technicolour models, which, otherwise, would contain a priori any theory exhibiting fermions with spontaneously broken chiral symmetry. For one, the space for electroweakly charged matter beyond what is already contained in the standard model is limited. 

Technicolour alone provides the masses of the weak gauge bosons. It does not explain where the masses of the standard model fermions $\psi$ come from. In the standard model this is accomplished by the Yukawa couplings to the Higgs $H$, $\sim\langle H\rangle\bar{\psi}\psi$. In the technicolour context the chiral condensate $\langle\bar{Q}Q\rangle$ of the techniquarks $Q$ replaces the non-zero Higgs condensate, and we could consider the operator $\sim\langle\bar{Q}Q\rangle\bar{\psi}\psi$. It is not renormalisable and in order to have a renormalisable theory it must be induced at energies above the technicolour chiral symmetry breaking scale. From Fermi's theory \cite{Fermi:1934} and its UV completion in form of the Glashow-Weinberg-Salam model \cite{Glashow:1961tr}, we know that these operators arise from the exchange of massive gauge bosons that acquire their mass at a higher energy scale (the electroweak scale) than the process under consideration (e.g., $\beta$ decay). Applying this approach in the technicolour context is known as extended technicolour \cite{Dimopoulos:1979es}. 
In those models, in general, the required four-fermion operator $\sim\bar{Q}Q\bar{\psi}\psi/\Lambda_\mathrm{ETC}^2$, where $\Lambda_\mathrm{ETC}$ denotes the scale at which the gauge bosons acquire their mass, are accompanied by two other types, $\sim\bar{Q}Q\bar{Q}^\prime Q^\prime/\Lambda_\mathrm{ETC}^2$ and $\sim\bar{\psi}\psi\bar{\psi}^\prime\psi^\prime/\Lambda_\mathrm{ETC}^2$. The former, for example, contribute to the masses of potentially present pseudo--Nambu-Goldstone modes. The latter induce flavour-changing neutral currents, which are known to be very small in nature. They cannot be reduced enough by simply choosing the scale $\Lambda_\mathrm{ETC}$ sufficiently big, as otherwise the top quark mass cannot be reached. (In view of the flavour-changing neutral currents, which have to be avoided, in the technicolour context the masses of all the standard model fermions apart from the top quark are natural, while in the context of the standard model order one Yukawa couplings would appear natural and, as a consequence, only the top mass.) In the standard model the flavour-changing neutral currents are suppressed by the Glashow-Iliopoulos-Maiani mechanism \cite{Glashow:1970gm}. In connection to extended technicolour there are several ways of suppressing the flavour-changing neutral currents, some of which will be mentioned below and which allow for selecting viable models for dynamical electroweak symmetry breaking. 

The details of the construction depend crucially on whether the techniquarks carry ordinary (QCD) colour or not. In case they do, we can unify the generation index of the standard model fermions with the technicolour index, $\mathcal{G}_\mathrm{ETC}\supset\mathcal{G}_\mathrm{TC}\times\mathcal{G}_\mathrm{gen.}$ For three generations and the minimal number of two technicolours this could be realised as $SU(5)\supset SU(3)_\mathrm{gen.}\times SU(2)_\mathrm{TC}$ \cite{Ryttov:2010kc}.
The $SU(5)$ is broken down sequentially to $SU(3)_\mathrm{gen.}\times SU(2)_\mathrm{TC}$ at three different scales $\Lambda_\mathrm{ETC,1}>\Lambda_\mathrm{ETC,2}>\Lambda_\mathrm{ETC,3}$ to explain the generational mass hierarchy of the standard model fermions. Here the mass operator of the $j$-th generation contains the $j$-th scale, $\sim\bar{Q}Q\bar{\psi}_j\psi_j/\Lambda_{\mathrm{ETC},j}^2$. Flavour-changing--neutral-current transitions always occur between generations, and a detailed analysis of the operators by which they are induced shows that they are suppressed relative to the above na{\"\i}ve discussion by powers of $\Lambda_{\mathrm{ETC}<}/\Lambda_{\mathrm{ETC}>}$ \cite{Appelquist:2002me}, where $\Lambda_{\mathrm{ETC}<}$ ($\Lambda_{\mathrm{ETC}>}$) is the smaller (bigger) of the two scales.

Technicolour models in which the techniquarks do not carry ordinary colour are more frugal when it comes to adding matter: There are (up to) a factor of three less additional fermion species from the viewpoint of the electroweak interactions. When it comes to the generation of the quark masses, some of the gauge bosons of the extended-technicolour sector must carry quantum chromodynamics colour. Thus, in this case $SU(3)_\mathrm{QCD}$ must be embedded in the extended technicolour gauge group \cite{Appelquist:1996kp}, $\mathcal{G}_\mathrm{ETC}\supset SU(3)_\mathrm{QCD}\times\mathcal{G}_\mathrm{TC}\times\mathcal{G}_\mathrm{gen.}$ (One can even consider models in which the extended technicolour sector does not commute with the weak interactions, which is known as noncommuting extended technicolour \cite{ref:noncometc}.)\footnote{
In the context of strong electroweak symmetry breaking, at the effective field theory level frequently a mechanism providing the standard-model fermion masses through mixing with bound states is mentioned \cite{Kaplan:1991dc}. Originally, these bound states were thought to be technihadrons. In the first place, this necessitates techniquarks with ordinary colour in order to have bound states that can mix with the ordinary quarks. Furthermore, it constrains the gauge group and representations of the techniquarks under the technicolour gauge group, as the bound state must be a singlet under technicolour and in the fundamental representation of ordinary colour. Interestingly, the same effective operators arise naturally in extended technicolour as bound states of techniquarks with extended-technicolour gluons bound by technicolour gluons. In extended technicolour these bound states always have automatically the correct quantum numbers.}

In both of the above types of models an enhancement of the techniquark condensate alleviates the tension between the small flavour-changing neutral currents and the comparatively large top mass. This can be accomplished in so-called walking theories. Because of a sufficient content of screening fermions, they almost evolve into an infrared fixed point \cite{Appelquist:1986an}. It is only missed because the critical coupling for chiral symmetry breaking to occur is reached just before the would-be Caswell-Banks-Zaks fixed point \cite{Caswell:1974gg}. As a consequence, the $\beta$ function for the gauge coupling at relatively large coupling is close to zero, and thus, the gauge coupling can be strong already for a large range of scales before the chiral condensate finally forms.
This entails a large anomalous dimension $\gamma$ for the quark mass operator, which leads to a sizeable renormalisation of the quark condensate: Na{\"\i}vely, the condensate $\langle\bar{Q}Q\rangle$ is expected to be of the order of $\Lambda_\mathrm{TC}^3$, where $\Lambda_\mathrm{TC}$ stands for the fundamental scale of the technicolour sector, i.e., the analogue of $\Lambda_\mathrm{QCD}$ in quantum chromodynamics. The mass operator $\bar{Q}Q\bar{\psi}\psi/\Lambda_\mathrm{ETC}^2$, however, is generated at the scale $\Lambda_\mathrm{ETC}$, whereas the condensate is formed at the smaller scale $\Lambda_\mathrm{TC}$. As a consequence, the renormalisation of the condensate between these two scales must be taken into account, $\langle\bar{Q}Q\rangle_\mathrm{ETC}=\langle\bar{Q}Q\rangle_\mathrm{TC}\exp\int_{\Lambda_\mathrm{TC}}^{\Lambda_\mathrm{ETC}}\frac{d\mu}{\mu}\gamma(\mu)$. In a walking theory the anomalous dimension $\gamma$ is approximately constant between these scales and the condensate is enhanced by a factor of $(\Lambda_\mathrm{ETC}/\Lambda_\mathrm{TC})^\gamma$. 

From what was said above, walking theories require a sufficient amount of matter, while electroweak precision data prefers technicolour sectors with a small amount of matter. Walking with a relatively small amount of matter can be achieved by choosing the techniquarks in a different than the fundamental representation of the technicolour gauge group \cite{Lane:1989ej,Dietrich:2005jn}. (Gauging only some, i.e., at least two, of the techniquarks under the electroweak interactions permits to further reduce the contribution to the oblique parameters \cite{Kennedy:1988sn}, while allowing for walking dynamics. This is known as partially gauged technicolour \cite{Dietrich:2005jn}.) 

Next to the additional gauge interactions, the additional bound states of the technicolour sector can become relevant for the physics of ultrahigh-energy cosmic rays. The generically lightest bound states are the technipions. In models with minimal flavour symmetry there are only three pions, which become the longitudinal degrees of freedom of the weak gauge bosons. For larger chiral symmetries there are more pions, which from electroweak corrections obtain masses of the order of the Z mass \cite{Dashen:1969eg}.  
In the context of the present study, models based on matter in higher-dimensional representations posses additional relevant features. 
(See, e.g., Sect.~\ref{sec:bs}.)

Taking stock, for the present project, a relevant aspect is that in technicolour models with techniquarks that carry ordinary colour the technicolour sector couples to standard matter already through quantum chromodynamics, whereas in technicolour models with all fermions singlets of ordinary colour it couples to standard matter only through the extended-technicolour sector. 

~

There are also other realisations of the dynamical electroweak symmetry breaking paradigm. Quite similar to technicolour is topcolour \cite{Hill:1991at}, where the condensation of the top quark due to a strong gauge interaction is responsible for the breaking of the electroweak symmetry. (The idea that the condensation of the top could be responsible for electroweak symmetry breaking \cite{ref:topcondensation} predates its gauge completion.) In order to implement the topcolour mechanism the SU(3) of quantum chromodynamics is embedded  (at least) in SU(3)$\times$SU(3), where the first two quark generations are charged under one of the SU(3)-s and the third generation under the second. \{One could also think of using an additional U(1) \cite{ref:topcondensationU1}.\} At a scale $M$, this group is then broken down to the diagonal subgroup, which represents the SU(3) of quantum chromodynamics. This breaking induces a four-fermion interaction for the top quark. Once the coupling of this interaction reaches its critical value, the condensation of the top is triggered. For small $M\approx 1$TeV the top quark would have to have a mass of about 600GeV. Only for $M\approx 10^{15}$GeV it may be as light as 160GeV \cite{Hill:2002ap}. Low-scale topcolour ($M\lll 10^{15}$TeV) with the physical top mass can be achieved by combining it with a technicolour sector into topcolour assisted technicolour \cite{Hill:1991at,ref:TC2}. (This is a straightforward combination in the sense that in technicolour the extended technicolour sector generically induces all kinds of four-fermion interactions. Among them is also a four-top interaction. If it has the appropriate coupling strength both a technicolour and a top condensate can form.) Alternatively, by adding more matter a top seesaw \cite{ref:topseesaw} can be implemented to achieve a light enough top quark.

~

Furthermore, there is a class of models in which chiral and electroweak symmetry breaking do not occur simultaneously, but where the emergent Nambu-Goldstone fields supply the Higgs sector that is required for breaking the electroweak symmetry. This pseudo-Nambu-Goldstone Higgs has a compositeness scale that is decoupled from the electroweak symmetry breaking scale by virtue of the Goldstone theorem \cite{ref:goldstone}. By arguments of naturality it may, however, not be too large. Subsequently, in these approaches an appropriate Higgs potential must be generated for the Nambu-Goldstone modes, such that electroweak symmetry breaking can occur. One type of these approaches is known as composite Higgs models \cite{ref:composite,ref:compositeads}. (Nowadays, they are frequently analysed on AdS$_5$ \cite{ref:compositeads}, as are sometimes technicolour models \cite{ref:tcads}.) These models are characterised by a vacuum misalignment parameter $\xi=v^2/f^2$, where $v$ is the analogue of the Higgs condensate and $f$ represents the pion decay constant. $\xi\rightarrow 1$ is commonly called the technicolour limit, while $\xi\rightarrow 0$ is referred to as standard model limit, where the compositeness scale is much larger than the electroweak scale. 
Another set of these models is known as Little Higgs \cite{ref:littlehiggs}, in which particular care is taken to eliminate radiative corrections to the Higgs mass appearing at one loop. They feature yet more gauge bosons needed for generating the Higgs potential, which, in the lightest case, are heavier than 2TeV \cite{ref:littlehiggsrev}.


\section{Cosmic-ray--air interaction\label{sec:main}}

In the energy range of importance for dynamical electroweak symmetry breaking the cosmic rays are detected by observing the air showers they initiate. We must, hence, identify the principal contributions to the cross section of cosmic-ray with atmospheric particles. 
Let us start with hadronic cosmic-ray particles. (We will drop this assumption in Sect.~\ref{sec:bsm}.) Within the standard model their interaction with the atmosphere is dominated by quantum chromodynamics. Hence, the dominant contribution to the scattering by quantum chromodynamics sets the benchmark for all processes arising from extensions of the standard model.

\begin{figure}[t]\begin{center}
 \resizebox{!}{7.5cm}{\includegraphics{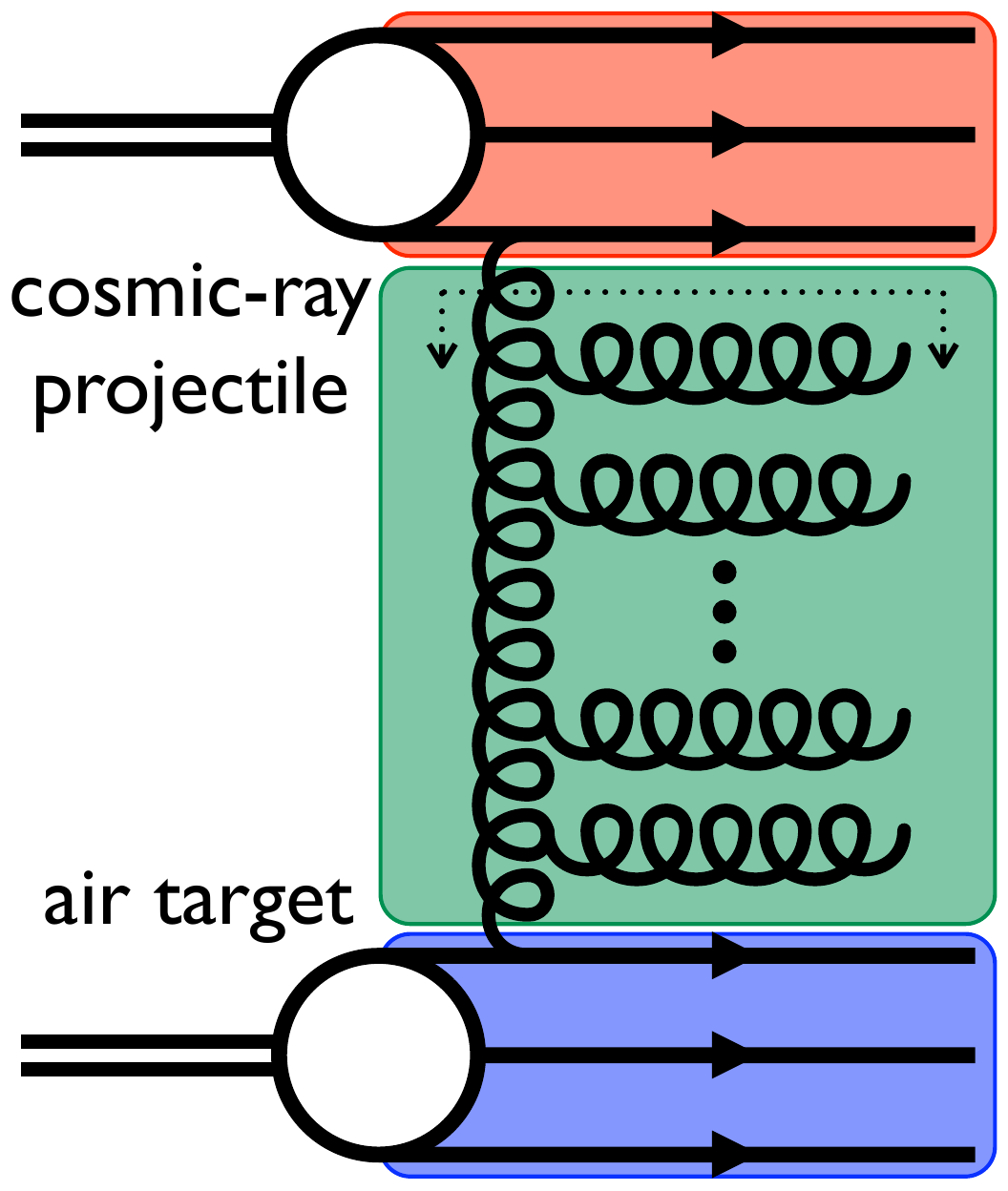}}
 \caption{In hadron-hadron scattering in ultrahigh-energy cosmic rays the leading parton from the projectile (top box, red, valence quarks shown, solid lines) probes the target (bottom box, blue) at very small momentum fraction. Therefore, the cross section is increased by the emission of additional gluons (curly lines, middle box, green).
 }
 \label{fig:qcd_bfkl}
 \end{center}\end{figure}

\begin{figure}[t]\begin{center}
 \resizebox{!}{7.5cm}{\includegraphics{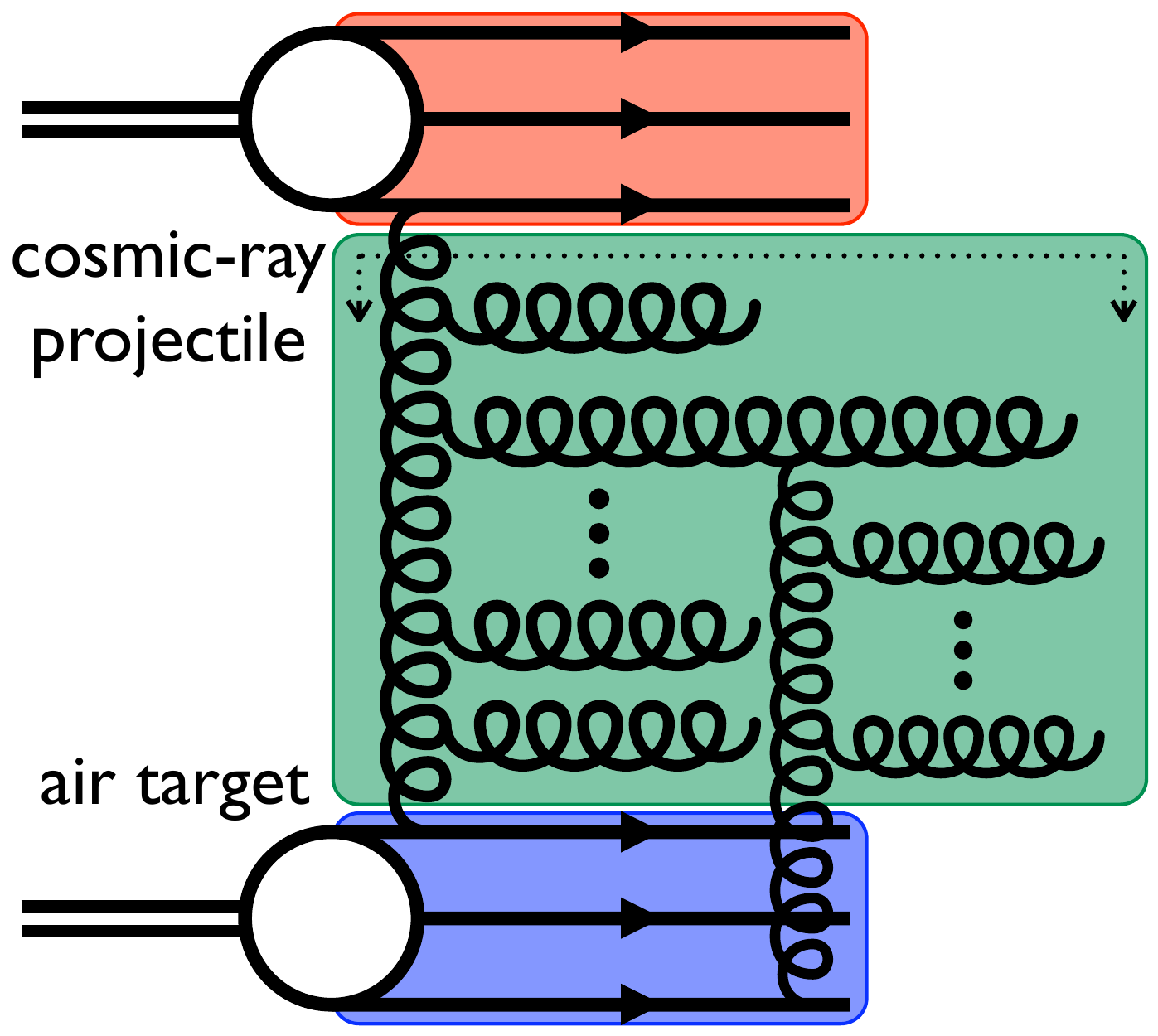}}
 \caption{Saturation. The gluon density and hence the growth of the cross section slows down once the density is so large that the loss term (example shown) compensates the effect from the gain term (the emission of an extra gluon).}
 \label{fig:qcd_sat}
 \end{center}\end{figure}

Since forward particles carry the largest fraction of the energy, they dictate the evolution of the air shower \cite{Drescher:2004sd}.\footnote{Due to Lorentz invariance, the computations could of course be carried out in any Lorentz frame, but the lab frame is the most intuitively clear one, as in it the primary target, the targets of the secondaries, and the detectors are (approximately) at rest. Lorentz invariance is manifest in Eq.~(\ref{eq:fixed}).}
At these large values of Feynman $x_F$ (the longitudinal momentum fraction) the partons in the incident cosmic-ray particle (the ``projectile'') are probed in the dilute region whereas the atmospheric particle (the ``target'') is practically at rest and appears dense, as its parton distribution is probed at a small momentum fraction $x$. The dominant scattering process for the production of forward particles that drive the evolution of the shower is the liberation of quarks from the cosmic-ray particle by scattering off the strong gluon fields in the atmospheric particle. 
The quark essentially punches through the atmospheric particle and receives some transverse momentum that is still negligible compared to its longitudinal momentum, which is why one treats the process in the eikonal approximation.
The gluon fields arise because at small momentum fraction $x$ the emission of additional gluons is enhanced by large logarithms $\ln(1/x)$, such that $\alpha_\mathrm{QCD}\ln(1/x)$ is no longer perturbatively small---despite of the gauge coupling $\alpha_\mathrm{QCD}$ being small at large virtualities due to asymptotic freedom---and the corresponding series must be resummed to all orders in $\alpha_\mathrm{QCD}\ln(1/x)$. (See Fig.~\ref{fig:qcd_bfkl}.)
%
%
This gives rise to the Balitsky-Fadin-Kuraev-Lipatov (BFKL) evolution equations \cite{Lipatov:1976zz}. They predict a rapid growth of the gluon density---accompanied by a corresponding growth of the cross section---at small momentum fractions $x$ or, equivalently, large energies at fixed virtuality. This rise has been observed at HERA to be $\propto x^{-0.3}$ at virtualities of more than (2GeV)$^2$ \cite{ref:growth}. The power-law behaviour can be reproduced by the leading-order BFKL equation. There, however, the exponent comes out too large. Incorporating the next-to-leading order \cite{Fadin:1998py}, which resums terms of order $\alpha_\mathrm{QCD}\times[\alpha_\mathrm{QCD}\ln(1/x)]^n$ leads to a growth rate that is more consistent with experiment \cite{Triantafyllopoulos:2002nz}. In this context, an important aspect of orders beyond the leading is the running of the gauge coupling. 
Be it leading order, next-to-leading order, fixed or running coupling, in the theoretical description laid out so far, the gluon density would keep growing ad infinitum for decreasing
momentum fraction.
At variance to the Dokshitzer-Gribov-Altarelli-Parisi (DGLAP) evolution \cite{Gribov:1972ri}, which resums large logarithms $\ln Q^2$ of virtuality, and where the transverse size of the gluons, i.e., $\sim Q^{-2}$, becomes smaller faster than their number density increases, in the case of changing rapidity $y=\ln(1/x)$ the transverse size is constant. As a consequence, gluons must start overlapping at some point and non-linear effects must become important. 
Loss terms become important (see Fig.~\ref{fig:qcd_sat}) and will catch up with the gain terms.
Furthermore, a ceaselessly growing gluon density would lead to a violation of the Froissart bound \cite{Froissart:1961ux} for the cross section and hence of unitarity. As a consequence, the increase of the gluon density should level off when it is of the order $\alpha^{-1}_\mathrm{QCD}$ \cite{Mueller:1999wm}. Coming back to the picture of overlapping gluons, for a given rapidity $y$ there must be a virtuality $Q_\mathrm{QCD}^2(y)$ below which the hadron/nucleus at a given impact parameter appears essentially black; the gluon density for smaller virtualities is saturated and cannot grow any further. The quantity $Q_\mathrm{QCD}^2(y)$ is known as saturation momentum, and can be interpreted as the density of gluons per unit of rapidity and per transverse area at saturation. 
%
Staying within the present overlap picture, the most parsimonious way of estimating it is to fix it at the point where the linear evolution equations would predict the scattering of a colour dipole of the size $Q^{-2}_\mathrm{QCD}$ to be unity. 
This bound can be studied in much greater detail, e.g., in the framework of the Balitsky-Kovchegov (BK) equation \cite{Balitsky:1995ub}, which also leads to the identification of geometric scaling at HERA \cite{Stasto:2000er} with travelling waves \cite{Munier:2003vc} in the BK approach. In fact, geometric scaling means that because of saturation the cross section for deep inelastic scattering, which, in general, can be a function independently of the virtuality $Q^2$ and the rapidity $y$ becomes a function only of the ratio of the virtuality and the saturation momentum, $Q^2/Q^2_\mathrm{QCD}(y)$, the latter of which is a function of the rapidity $y$.
As explained above, the fast quark punches through the atmospheric nucleus at a given impact parameter. Hence, the encountered colour charge density depends on the longitudinally integrated nuclear density profile $\rho(r)$, i.e., the nuclear thickness function $T(b)=\int dz~\rho(r=\sqrt{b^2+z^2})$. If we adopt the normalisation $\int d^2b~T(b)=1$ an additional factor of the total number of nucleons, i.e., the atomic mass number $A$ must appear. From there the expected number of participating nucleons $P$ is obtained by multiplying by the inelastic 
projectile-nucleon cross section $\sigma_N$ for the given centre-of-mass energy $\sqrt{s}$,
\be
P=A~\sigma_N~T .
\label{eq:part}
\ee
Finally, the initial condition for the evolution of the saturation momentum is given by \cite{Drescher:2004sd}
\be
Q_\mathrm{QCD}^2(0)/\Lambda_\mathrm{QCD}^2=(1+P)\ln(1+P),
\label{eq:imppar}
\ee
The evolution in rapidity of this initial condition using the leading order BFKL kernel with a constant coupling constant leads to \cite{Mueller:1999wm,ref:fixed}
\be
Q^2_\mathrm{QCD}/Q^2_\mathrm{QCD}(0)=e^{c\bar{\alpha}_\mathrm{QCD}y} ,
\label{eq:fixed}
\ee 
where $\bar{\alpha}_\mathrm{QCD}=\alpha_\mathrm{QCD}N_\mathrm{QCD}/\pi$, $c\approx 5$, $N_\mathrm{QCD}=3$ stands for the number of ordinary colours\footnote{
We have adopted the lab frame for our discussion. In a general frame, say with a rapidity $\eta$ relative to the lab frame, the afore-described evolution of the gauge-boson densities takes place in the target as well as in the projectile. The projectile, which before remained unevolved, picks up an exponential factor $e^{c\bar{\alpha}_\mathrm{QCD}\eta}$, while the evolution factor of the target is altered to $e^{c\bar{\alpha}_\mathrm{QCD}(y-\eta)}$. Thus, the product of the two factors remains unchanged, which makes the above considerations frame independent. In Fig.~\ref{fig:qcd_bfkl} the choice of a different frame corresponds to a shift of the dotted line in the direction of the projectile, i.e., to a different partition of the emitted gluons between target and projectile, respectively.
} 

As next ingredient we need the cross section of a fast parton from the cosmic-ray particle on the dense gluons of the atmospheric nucleus. Here, the McLerran-Venugopalan model \cite{McLerran:1993ni}, which is formulated based on stochastic gluon fields, allows to determine an analytically closed expression \cite{Dumitru:2002qt} 
\be
d\sigma_\mathrm{tot}^\mathrm{QCD}/d^2b=2(1-e^{-\pi^2Q_\mathrm{QCD}^2/N_c\Lambda^2_\mathrm{QCD}}).
\label{eq:xsec}
\ee
(The computation involves solving the equation of motion for the probe in a classical background field, a widely used approach in this subject area \cite{ref:class}, and a subsequent averaging over configurations.)
The nuclear density function be given by the Woods-Saxon distribution \cite{Woods:1954zz}
\be
\rho_\mathrm{WS}(r)=n[1+e^{(r-R)/a}]^{-1},
\ee
with the normalisation constant
\bea
3n^{-1}
&=&
4\pi a^3[(R/a)^3+\pi^2 R/a-6\mathrm{Li}(3,-e^{-R/a})]
=\nonumber\\&=&
4\pi(R^3+R\pi a^2)+O(e^{-R/a}),
\eea
where Li stands for the polylogarithm, $a\approx 0.5$ fm for the nuclear surface thickness, and $R=(1.19A^{1/3}-1.61A^{-1/3})$fm \cite{Eskola:1995zt} for the nuclear radius. When encountering the atmosphere, the cosmic-ray particle is most likely to hit a nitrogen ($A=14$) or oxygen ($A=16$) nucleus and hence $R\approx 2.2$ fm. The corresponding nuclear thickness function at large impact parameter---where it will be most important below---is given by
\be
T(b)\approx\sqrt{2\pi ab}ne^{(R-b)/a}.
\label{eq:thick}
\ee
Integration of Eq.~(\ref{eq:xsec}) over the impact parameter $b$ yields for large total rapidity $y$,
\be
\sigma_\mathrm{tot}^\mathrm{QCD}
\approx
2\pi(\tilde R_\mathrm{QCD}+ac\bar{\alpha}_\mathrm{QCD}y)^2 ,
\label{xsectotfix}
\ee
which saturates the Froissart bound $\sigma_\mathrm{tot}\propto y^2$. Such a growth law is also consistent with the latest findings of the TOTEM collaboration up to $\sqrt{s}= 8$ TeV \cite{Antchev:2013paa}. Here
\be
\tilde R_\mathrm{QCD}=R+a\ln\frac{\pi^3An\sqrt{2\pi aR}}{N_c\Lambda_\mathrm{QCD}^2} .
\label{eq:tilder_qcd}
\ee
We have used Eq.~(\ref{eq:imppar}) to leading order in small $P$, Eq.~(\ref{eq:fixed}) and Eq.~(\ref{eq:thick}) with $b$ in the square root approximated by $R$ as well as the approximation $1-\exp(-e^{(b_0-b)/a})\approx\theta(b_0-b)$. Moreover, we have taken $\pi/\Lambda_\mathrm{QCD}^2$ as lower estimate for $\sigma_N$, which actually grows larger than that at high energies. 

As mentioned before there are effects beyond the leading order like the running of the coupling. Incorporating the latter
\be
\bar{\alpha}_\mathrm{QCD}(Q^2)=\bar{\beta}_\mathrm{QCD}/\ln(Q^2/\Lambda_\mathrm{QCD}^2) ,
\label{eq:rc}
\ee
the evolution slows down gradually and one gets to leading order in large rapidities \cite{ref:running}
\be
Q^2_\mathrm{QCD}\propto e^{\sqrt{2\bar{\beta}_\mathrm{QCD}cy}},
\ee
where $\bar{\beta}_\mathrm{QCD}=12N_c/(11N_c-2N_f^\mathrm{QCD})$, $N_f^\mathrm{QCD}$ stands for the number of quarks, and the momentum scale is set by the overall saturation momentum or the saturation momentum in an individual dipole, depending on which of the two is greater. 
Then the analogue of Eq.~(\ref{xsectotfix}) reads 
\be
\sigma_\mathrm{tot}^\mathrm{QCD}
\approx
2\pi(\tilde R_\mathrm{QCD}+a\sqrt{2c\bar{\beta}_\mathrm{QCD}y})^2 .
\label{eq:sigtotrun}
\ee


\subsection{Gauge interactions beyond the standard model}

Extensions of the standard model that break the electroweak symmetry dynamically include additional gauge interactions. Here we are going to investigate, whether these could lead to comparable contributions to the interaction of hadronic cosmic rays with the atmosphere. In particular, we study whether a BFKL-like evolution could achieve this. In hadron-hadron scattering at cosmic-ray energies quantum chromodynamics is already in the saturation regime and leads to a sizeable cross section. Hence, in this system the onset of the BFKL evolution in additional gauge sectors is unlikely to show up as a prominent feature in the cosmic-ray spectrum. Thus, the question must be, whether the new gauge bosons also manage to form a dense system and whether this system contributes significantly to the cross section. The obstacles to be overcome by the evolution are the suppression factors that arise from the coupling to the valence quarks in the projectile and the target, respectively. The coupling in the target will lead to a reduction in initial density, while the coupling in the projectile leads to an overall suppression of the result even for asymptotically large energies.


\subsubsection{With ordinary colour}
 
 \begin{figure}[t]\begin{center}
 \resizebox{!}{8cm}{\includegraphics{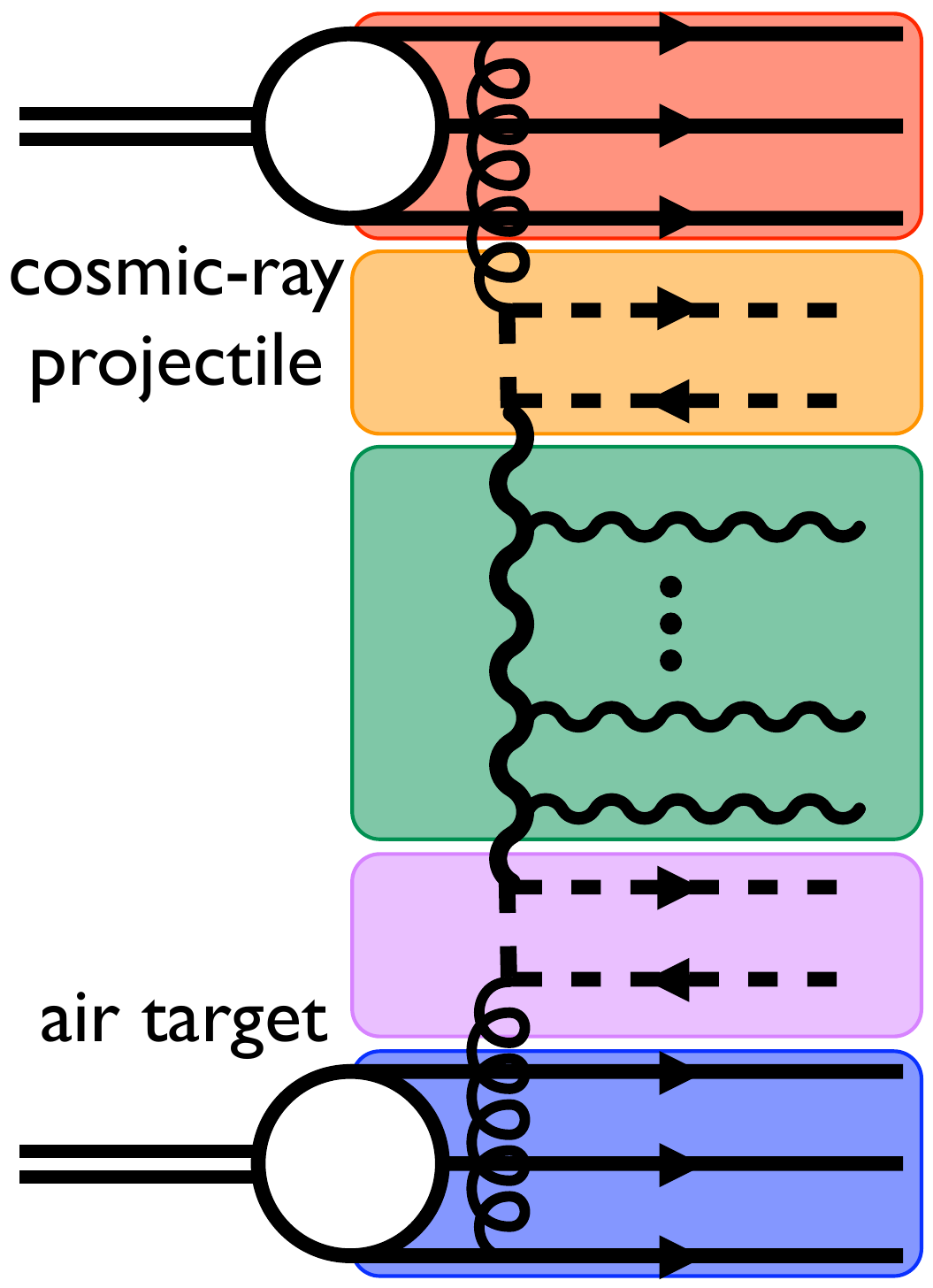}}
 \caption{Contribution from technicolour to the hadron-hadron [top (red) and bottom (blue) boxes, valence quarks shown, solid lines] cross section. The technigluons (wavy lines, middle box, green) are exchanged between techniquarks (dashed lines, second and fourth boxes, orange and lavender) that in this example also carry ordinary colour and can thus be produced from an ordinary gluon.}
 \label{fig:tc1}
 \end{center}\end{figure}
%
Let us start with a technicolour model with some techniquarks that also carry ordinary colour. (See Fig.~\ref{fig:tc1}.) (This is not the case in all technicolour models, but their specific contribution offers a suitable starting point, and we can draw on the corresponding results in what follows.) These techniquarks can arise from the splitting of a gluon. [Thus, this process does not involve the extended technicolour sector. (Additional) contributions from that sector are discussed below.] Therefore, at large rapidity $y$ and virtuality $Q^2$, their density $x\mathcal{Q}(x,Q^2)$ inside hadrons and nuclei should be like that of heavy quarks.  
It is governed by the DGLAP evolution equation \cite{Gribov:1972ri} for the emission of a fermion-antifermion pair from a gluon,
\be
\frac{\partial\mathcal{Q}(x,Q^2)}{\partial\ln Q^2}=2T_RN_f\frac{\alpha_\mathrm{QCD}}{4\pi}\int_x^1\frac{dz}{z}[z^2+(1-z)^2]\mathcal{G}(\frac{x}{z},Q^2) ,
\ee
where we summed over $N_f$ techniflavours, with trace normalisation factor $T_R$ and representation $R$ under ordinary colour.
Taking as initial condition that at the fundamental scale $\Lambda_\mathrm{TC}$ the density of this fermion species is zero, the solution of this equation is approximated well by \cite{Olness:1987ep}
\be
\mathcal{Q}(x,Q^2)\approx 2T_RN_f\frac{\alpha_\mathrm{QCD}}{4\pi}\ln\frac{Q^2}{\Lambda_\mathrm{TC}^2}\mathcal{G}(x,Q^2)
\ee
for small enough momentum fraction $x$. Here, neglecting the scale dependence of the product of the coupling and the gluon distribution, gluon bremsstrahlung and taking the $z$ integral to be dominated by small $x/z$ because of the rapid growth of the gluon distribution for small momentum fraction conspire to yield this result. Taking into account the integration volume by multiplication by $(1-x)$ improves the approximation for cases where $x\slashed\ll1$. 
In the dilute regime, i.e., for $Q^2\approx\Lambda_\mathrm{TC}^2\gg Q_\mathrm{QCD}^2$ 
the gluon density is given by the Weizs\"acker-Williams radiation \cite{von Weizsacker:1934sx} off $A\times N_c$ separate quarks in the target, 
\be
x\mathcal{G}(x,Q^2)=\frac{\alpha_\mathrm{QCD}AN_cC_F^\mathrm{QCD}}{\pi}\ln\frac{Q^2}{\Lambda_\mathrm{QCD}^2} .
\label{eq:gluedens}
\ee
Consequently,
\be
x\mathcal{Q}_A(x,Q^2)
\approx 
\frac{T_RN_f\bar{\beta}_\mathrm{QCD}^2AC_F^\mathrm{QCD}}{2N_c}\frac{\ln\frac{Q^2}{\Lambda_\mathrm{TC}^2}}{\ln\frac{Q^2}{\Lambda_\mathrm{QCD}^2}} ,
\label{eq:diptar}
\ee
where we have made use of Eq.~(\ref{eq:rc}). Therefore, the number of participants, the analogue of Eq.~(\ref{eq:part}), is given by
\be
P_\mathrm{TC}\approx x_t\mathcal{Q}_A\frac{\pi}{\Lambda^2_\mathrm{TC}}T(b).
\ee
For a proton projectile the dipole density is obtained by setting $A\rightarrow1$ in Eq.~(\ref{eq:diptar}), and thus the analogue of the cross section (\ref{eq:xsec}) at fixed $x_p$ and $x_t$ reads
\be
d\sigma_\mathrm{tot}^\mathrm{TC}/d^2b\approx x_p\mathcal{Q}_12(1-e^{-\pi^2Q_\mathrm{TC}^2/N_\mathrm{TC}\Lambda^2_\mathrm{TC}}).
\ee
Carrying out the integration over the impact parameter, one finds to leading order in large rapidities
\be
\sigma_\mathrm{tot}^\mathrm{TC}
\approx
\#2\pi(\tilde R_\mathrm{TC}+ac\bar{\alpha}_\mathrm{TC}y
)^2,
\label{eq:sigmatottc}
\ee
where
\be
\#=\frac{T_RN_f\bar{\beta}_\mathrm{QCD}^2C_F^\mathrm{QCD}}{2N_c}\frac{\ln\frac{Q^2}{\Lambda_\mathrm{TC}^2}}{\ln\frac{Q^2}{\Lambda_\mathrm{QCD}^2}} 
\label{eq:hash}
\ee
and
\be
\tilde R_\mathrm{TC} = R+a\ln\frac{A\#\pi^3\sqrt{2\pi aR}n}{N_\mathrm{TC}\Lambda_\mathrm{TC}^2} .
\label{eq:tilder}
\ee
For two fundamental flavours with two technicolours, $\Lambda_\mathrm{TC}=10^3\Lambda_\mathrm{QCD}$, and $\ln(Q^2/\Lambda_\mathrm{TC}^2)=1$ the expression (radius) in Eq.~(\ref{eq:sigmatottc}) exceeds $R-a$ for $c\bar{\alpha}_\mathrm{TC}y\gtrsim 13$, which marks the onset of saturation in the technigluon sector. The total rapidity at the GZK cutoff is $y\approx26$; hence, the above criterion corresponds to the value $c\bar{\alpha}_\mathrm{TC}\approx 0.5$, i.e., e.g., $\bar{\alpha}_\mathrm{TC}\approx 0.2$ and $c\approx 2.5$.
In the case of a running coupling, the result reads
\be
\sigma_\mathrm{tot}^\mathrm{TC}
\approx
\#2\pi(\tilde R_\mathrm{TC}+a\sqrt{2c\bar{\beta}_\mathrm{TC}y})^2,
\label{eq:sigmatottcrun}
\ee
which leads to the saturation criterion $cy\gtrsim 66$ corresponding to $c\approx 2.5$.


\subsubsection{Walking dynamics}

As touched upon in the introduction, a walking coupling is a desired feature for a technicolour theory. A walking coupling is approximately fixed over a certain range of scales and only starts decreasing sizeably once the system is analysed at yet higher scales. This could be taken into account by implementing the corresponding two-loop $\beta$ function in the vicinity of its fixed point when determining the saturation momentum. 
Approximating it by using the above results, the growth factor of the saturation momentum would take the form
\be
Q^2_\mathrm{TC}\approx Q^2_\mathrm{TC}(0)e^{c\bar{\alpha}y_\mathrm{walk}+\sqrt{c\bar{\beta}(y-y_\mathrm{walk})}} ,
\ee
where $y_\mathrm{walk}$ denotes the rapidity interval for which the system stays in the walking regime, $\bar{\alpha}$ the coupling in the vicinity of the fixed point, and $\bar{\beta}$ the effective value of the $\beta$ function once the coupling starts decreasing again. $y_\mathrm{walk}$ is chosen such, that it takes the initial value $Q^2_\mathrm{TC}(0)$ of the saturation momentum to the scale $\Lambda^2_\mathrm{ETC}$ where the running sets in. Hence, we could write equivalently
\be
Q^2_\mathrm{TC}\approx\Lambda^2_\mathrm{ETC}e^{\sqrt{c\bar{\beta}\{y-\ln[\Lambda^2_\mathrm{ETC}/Q^2_\mathrm{TC}(0)]/(c\bar{\alpha})\}}} .
\ee


\subsubsection{So far}

Taking stock, even for only two techniflavours the available rapidity allows the additional gauge sector to reach saturation at half the theoretical leading-order value of the exponent $c$. At the same time QCD has a head start, due to its lower fundamental scale. Let us consider the relative contributions to the total cross section
\be
\frac{\sigma^\mathrm{TC}_\mathrm{tot}}{\sigma^\mathrm{QCD}_\mathrm{tot}}
=
\#\bigg(\frac{\tilde R_\mathrm{TC}+af_\mathrm{TC}(y)}{\tilde R_\mathrm{QCD}+af_\mathrm{QCD}(y)}\bigg)^2,
\ee
where $f(y)=c\bar\alpha y$ for fixed coupling and $f(y)=\sqrt{2c\bar\beta y}$ for running coupling.

\begin{figure}[t]\begin{center}
 \resizebox{7.5cm}{!}{\includegraphics{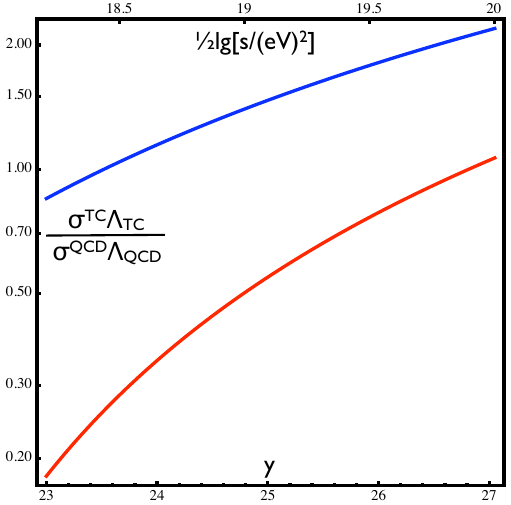}}
 \caption{Estimate for the expected value for the produced transverse momentum by a technicolour sector divided by the contribution from quantum chromodynamics for $c\bar{\alpha}_\mathrm{TC}=0.5=c\bar{\alpha}_\mathrm{QCD}$ at $N_f=2$ (red,bottom) and $N_f=8$ (blue, top), respectively, plotted versus rapidity $y$ and the centre-of-mass energy $\sqrt{s}$.}
 \label{fig:pt}
 \end{center}\end{figure}

At any given energy/rapidity the technigluon density should grow faster than the gluon density, as the fundamental scale of technicolour is much higher than that of quantum chromodynamics, which delays the asymptotic freedom. Additionally, in the typically highly flavoured technicolour theories, the running of the coupling is slower than in QCD. Hence, assuming equal evolution speeds $f_\mathrm{QCD}(y)\approx f_\mathrm{TC}(y)$ when analysing the above ratio yields a rather conservative estimate. Under these assumptions, at the onset of saturation,
\be
\frac{\sigma^\mathrm{TC}_\mathrm{tot}}{\sigma^\mathrm{QCD}_\mathrm{tot}}
>
\#\bigg(\frac{R-a}{\tilde R_\mathrm{QCD}-\tilde R_\mathrm{TC}+R-a}\bigg)^2
\approx
7\times 10^{-4}\frac{N_f}{2}.
\ee
Hence, for the above benchmark values the contribution from technicolour is of the order of a thousandth of that from quantum chromodynamics. Looking at the expected value for the transferred transverse momentum, at these high energies the two contributions are of the same order of magnitude, because the technicolour scale is about a thousand times that of quantum chromodynamics. Moreover, the ratio keeps growing for increasing rapidities even from the conservative viewpoint, i.e., with equal evolution rates. (See Fig.~\ref{fig:pt}.) At asymptotically high energies/rapidities the ratio would converge to $\#$, i.e., the reduced coupling of the technigluon ladder to the colliding hadrons, and otherwise would be enhanced by a factor of, e.g., $\bar{\alpha}_\mathrm{TC}^2/\bar{\alpha}_\mathrm{QCD}^2$ in the fixed coupling case. Because of the largely different fundamental scales and different final states there is no interference between the two sectors, and the expected values for the transverse momentum simply add up. 
%

The rare but very hard contributions from the technicolour sector are, however, masked by rare hard QCD processes like the emission of an additional top-antitop pair. In that case, relative to the process depicted in Fig.~\ref{fig:tc1}, the couplings of the top quarks to the gluons bring the cross section down by two factors of $\alpha_\mathrm{QCD}(2m_t)\approx 0.1$. Moreover, the larger momentum scale $2m_t\gg\Lambda_\mathrm{QCD}$ requires using the same reduced value for the coupling in the gluon ladder, which leads to an additional reduction. 
Overall the QCD cross section with an additional top antitop pair is of the same order of as the technicolour cross section. As the momentum scale of the QCD process is now about as big as that of the technicolour process, the expected values for the produced transverse momentum are comparable as well.

Hence, technicolour likely does not show up in hadron-hadron events as a striking change of the momentum structure. Still, the shower progression is different. The QCD gluons split into quarks, which ultimately hadronise. Analogously, the technigluons will first split into techniquarks. Depending on the details of the realisation of the technicolour model, these techniquarks either hadronise first into technihadrons or decay through the weak or extended-technicolour interactions already before. They can also form (meta)stable technihadrons. In both cases the free electric charge of the (techni)quarks and/or (techni)hadrons will lead to an electromagnetic shower. Also in both cases the generically lightest and thus most abundant (techni)hadrons are the (techni)pions. Three of the technipions are the longitudinal degrees of freedom of the weak gauge bosons, with their known decay products.\footnote{Moreover, technibaryon number is preserved (perturbatively) by the standard model and the technicolour interactions, which stabilises the lightest technibaryons against rapid decay. (In some technicolour models these can be additional technipions.)}
Therefore, the presence of technicolour could lead to a proliferation of weak gauge bosons and their decay products especially at the hard end of the spectrum. Whether such signals can be disentangled from the background in rare high-energy events is, however, a different question and must be left for later work.


\subsubsection{Within the standard model}

We should also discuss whether the SU(2)$_L$ of the electroweak interactions, an integral part of the standard model, could achieve something similar completely on its own. For this purpose the hadrons can be interpreted as dipoles right away albeit as rather dilute ones, as their scale is set by $\Lambda_\mathrm{QCD}$ whereas the SU(2)$_L$ can only be considered in this context for scales much larger than the Z mass. Because of the smallness of the coupling constant $\bar{\alpha}_\mathrm{L}\approx1/46$, the evolution parameter $c\bar{\alpha}_\mathrm{L}Y$ is not sufficient for the hadrons to become densely populated by these gauge bosons even at the GZK cutoff. 

~

To the contrary, if we use the weak gauge bosons as portal from the valence quarks to technidipoles instead of the gluon portal, relative to the latter contribution we pick up an additional factor of
\be
\frac
{\bar{\alpha}_\mathrm{L}^2}
{\frac{\bar{\beta}_\mathrm{QCD}^2}{\ln^2\frac{Q^2}{\Lambda_\mathrm{QCD}^2}}}
\frac{\ln\frac{Q^2}{\Lambda_\mathrm{EW}^2}}{\ln\frac{Q^2}{\Lambda_\mathrm{QCD}^2}}
=
\frac
{\bar{\alpha}_\mathrm{L}^2}
{\bar{\beta}_\mathrm{QCD}^2}
\ln\frac{Q^2}{\Lambda_\mathrm{EW}^2}\ln\frac{Q^2}{\Lambda_\mathrm{QCD}^2}
\approx
O(10^{-3})
\ee
(times some colour factors), which enters in the overall prefactor of the cross section and the initial condition for the technigluon density and makes this contribution strongly subleading. The first fraction on the left-hand side accounts for the different couplings, the second for the difference in phase space. For the massless photon there would still be a logarithmic enhancement 
$
\ln(Q^2/\Lambda_\mathrm{EW}^2)\rightarrow\ln(Q^2/m_u^2)\approx20\ln(Q^2/\Lambda_\mathrm{EW}^2),
$
which still cannot compensate for the powers of the weak interaction coupling.


\subsubsection{Without ordinary colour}
 
 \begin{figure}[t]\begin{center}
 \resizebox{!}{12cm}{\includegraphics{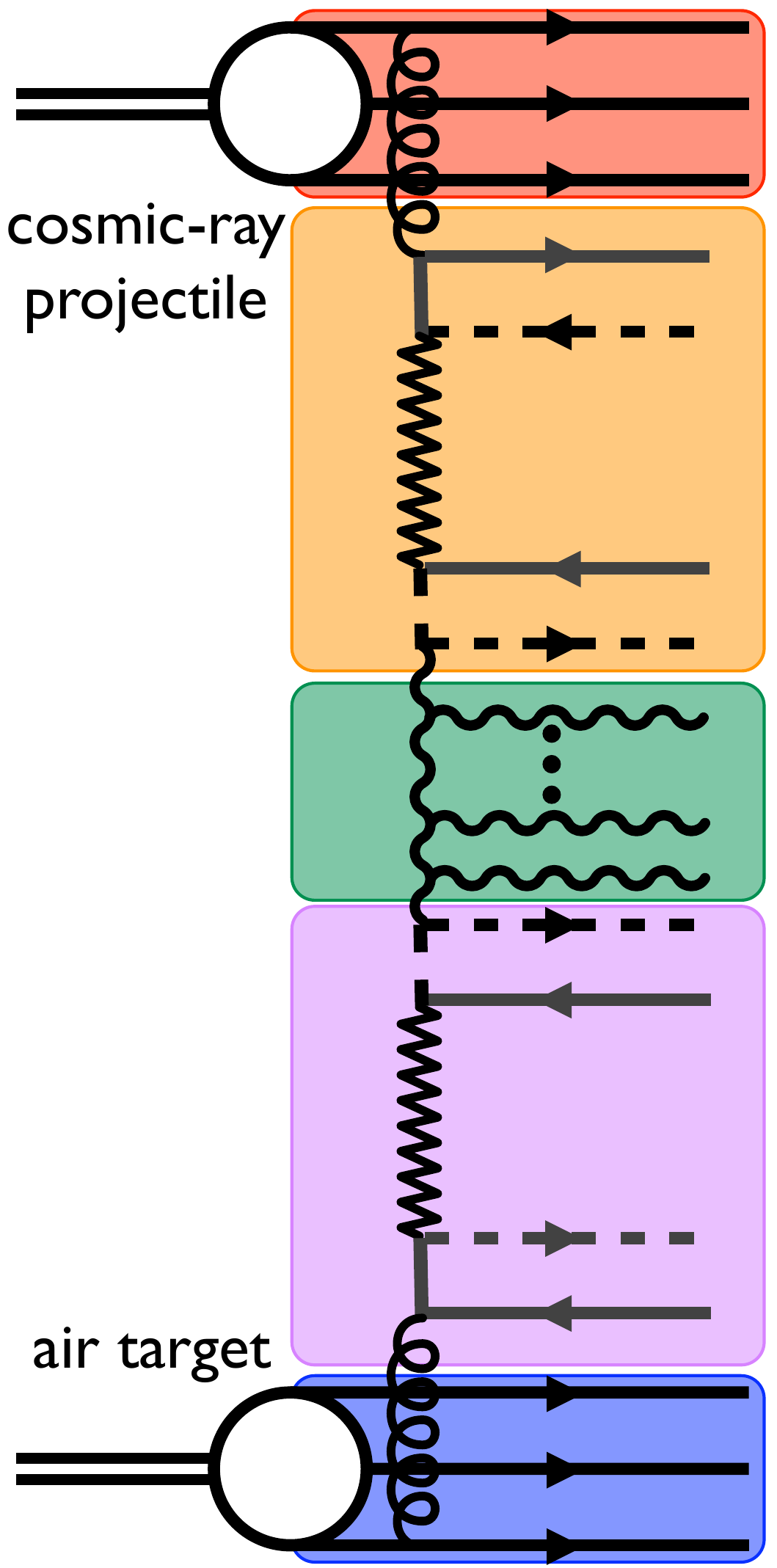}}
 \caption{Contribution from technicolour to the hadron-hadron [top (red) and bottom (blue) boxes, valence quarks shown, solid lines] cross section. The technigluons (wavy lines, middle box, green) are exchanged between techniquarks (dashed lines, second and fourth boxes, orange and lavender; the zigzag lines represent the extended technigluons), which for this contribution may or may not carry ordinary colour. They are linked to the hadrons through the extended-technicolour sector, the analogue of the top Yukawa coupling. For techniquarks that carry ordinary colour all combinations of gluon portals (See Fig.\ref{fig:tc1}.) with top-quark--extended-technicolour portals at the target and projectile sides, respectively, contribute.}
 \label{fig:tc2}
 \end{center}\end{figure}
%
After studying technicolour with techniquarks that are charged under ordinary colour, we continue for theories in which they are not. As pointed out in the introduction, these theories require less additional matter from the viewpoint of the electroweak interactions and are thus easier to accommodate within the bounds set by electroweak precision data. In these theories an ordinary gluon cannot emit a techniquark-antitechniquark pair. Thus this gateway is barred. The gauge bosons of the extended technicolour sector, however, must also carry ordinary colour in order to be able to give the ordinary quarks their mass. Hence, a valence quark in the hadrons can emit such an extended technigluon. (The entire consideration also goes through for a scalar extended-technicolour mechanism.) This would convert the quark to a techniquark, and we would have a technicolour dipole between the techniquark and the extended technigluon. The extended technigluon, however, is very heavy (for coupling to the first generation of the order $10^3$TeV) and we cannot put it on or even close to the mass shell with $\sqrt{s}$ about $10^2$TeV in the vicinity of the GZK cutoff. Therefore, we must consider a highly virtual extended technigluon that splits immediately into a quark-antitechniquark pair. (See Fig.~\ref{fig:tc2}.) This amounts to considering a four-fermion interaction involving two ordinary quarks and two techniquarks with the coupling strength $m_{u,d}/\langle\bar{T}T\rangle$, where $\langle\bar{T}T\rangle$ is the techniquark condensate, which is of the order of $\Lambda_\mathrm{TC}^3\approx0.1\mathrm{TeV}^3$ for running theories and larger for walking ones, while $m_{u,d}\approx 10^{-5}$TeV. Consequently, after squaring the amplitude we must overcome a factor of $(2\pi^3)^{-1}10^{-8}Q^4/\mathrm{TeV}^4$, where we have used the virtuality $Q^2$ to represent the momentum scale, and multiplying by $(2\pi^3)^{-1}$ accounts for the phase space for a $1\rightarrow 3$ process with two unresolved particles. To bring this factor anywhere close to unity requires again much more energy than is available even at the GZK cutoff.

\begin{figure}[t!]\begin{center}
 \resizebox{7.5cm}{!}{\includegraphics{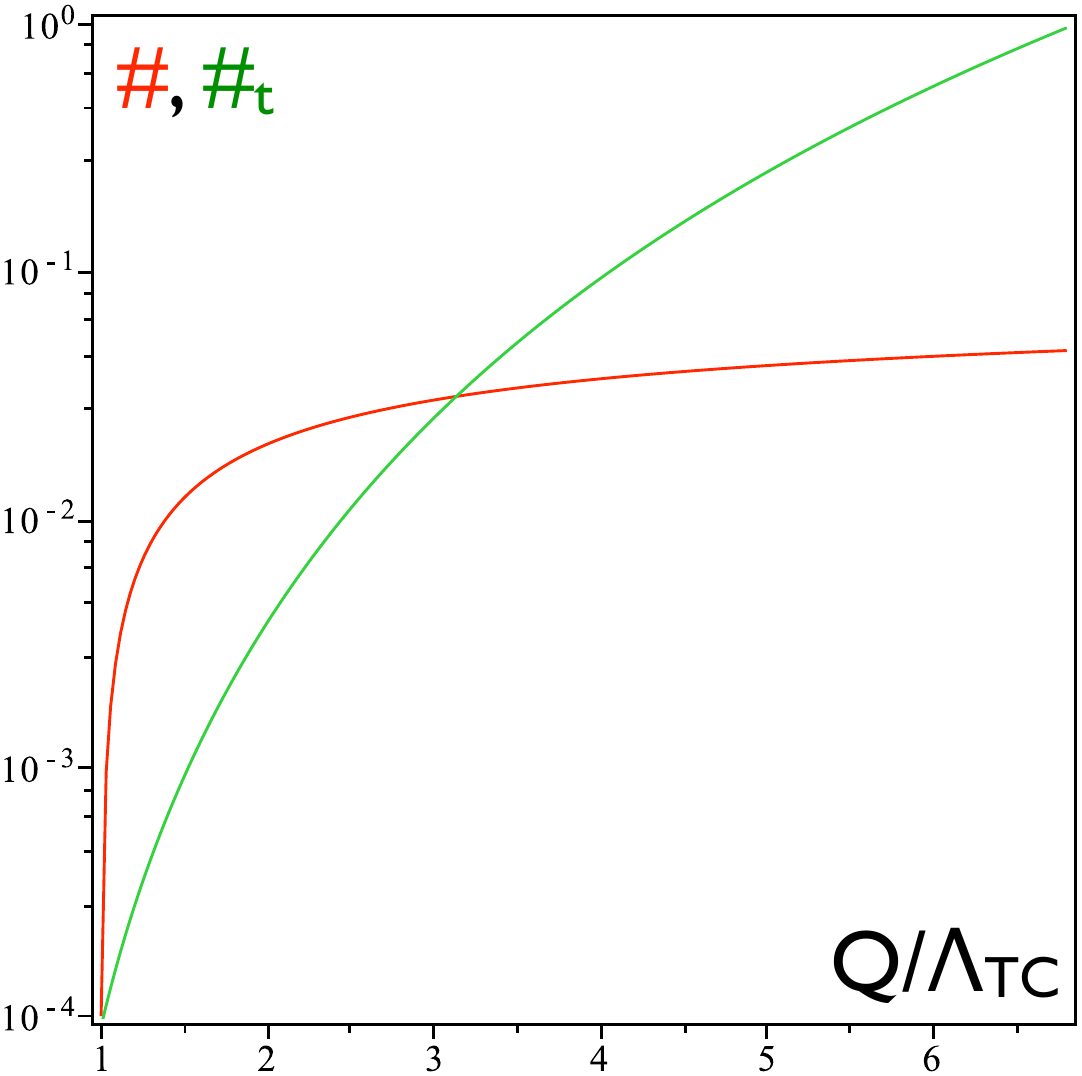}}
 \caption{Suppression factors $\#$ [Eq.~(\ref{eq:hash}), flatter (on the right-hand side) curve, red] and $\#_t$ [Eq.~(\ref{eq:hasht}), steeper curve, green] as functions of the scale $Q$ relative to the fundamental scale $\Lambda_\mathrm{TC}$ of technicolour for the benchmark values.}
 \label{fig:hash}
 \end{center}\end{figure}

The suppression is so overwhelming because of the smallness of the masses of the valence quarks compared to the technicolour scale (the equivalent of a small Yukawa coupling). Hence, we should take a look at contributions involving the top quark. (See Fig.~\ref{fig:tc2}.) Its distribution in the hadron is given by Eq.~(\ref{eq:diptar}) with the replacement $\Lambda_\mathrm{TC}\rightarrow m_t$,
\be
x\mathcal{T}_A(x,Q^2)
\approx 
\frac{T_RN_f\bar{\beta}_\mathrm{QCD}^2AC_F^\mathrm{QCD}}{2N_c}\frac{\ln\frac{Q^2}{m_t^2}}{\ln\frac{Q^2}{\Lambda_\mathrm{QCD}^2}} .
\label{eq:diptop}
\ee
If we couple the previous four-fermion interaction to the top or antitop, we only get an additional suppression of the order $(2\pi^3)^{-1}m_t^2/\Lambda_\mathrm{TC}^6$. The extended-technicolour scale for the extended technigluons coupling to the third generation of the standard model fermions is a few TeV. Hence, in detail depending on the value of the extended technicolour coupling, the mass of the extended technigluon is somewhat smaller than that value. Below that mass scale, we find the analogue of Eq.~(\ref{eq:sigmatottc}), but with
\be
\#_\mathrm{t}=\frac{m_t^2Q^4}{2\pi^3\Lambda_\mathrm{TC}^6}\frac{T_RN_f\bar{\beta}_\mathrm{QCD}^2C_F^\mathrm{QCD}}{2N_c}\frac{\ln\frac{Q^2}{m_t^2}}{\ln\frac{Q^2}{\Lambda_\mathrm{QCD}^2}}
\label{eq:hasht}
\ee
instead of $\#$, i.e.,
\be
\sigma_\mathrm{tot}^\mathrm{TCt}
\approx
\#_\mathrm{t}2\pi(\tilde R_\mathrm{TCt}+ac\bar{\alpha}_\mathrm{TC}y)^2 ,
\ee
where
\be
\tilde R_\mathrm{TCt}=R+
a\ln\frac{A\#_\mathrm{t}\pi^3\sqrt{2\pi aR}n}{N_\mathrm{TC}\Lambda_\mathrm{TC}^2} .
\ee
On this occasion we ignore additional combinatorial factors on the extended-technicolour side, involving, e.g., the rank of the gauge group etc. For the same benchmark values as above, but for $Q^2\approx(3.14\Lambda_\mathrm{TC})^2$ instead of $Q^2\approx2.72\Lambda_\mathrm{TC}^2$, $\#$ and $\#_t$ are equal and so are the two total cross sections. For smaller values of $Q^2$ we have $\#>\#_t$, while for larger values $\#_t>\#$. Hence, in the typical range of parameters the two contributions are comparable. (See Fig.~\ref{fig:hash}.)
%

If, however, the virtuality $Q^2$ should reach the mass of the lightest extended technigluons, which are involved in linking the top quark to the technigluons, $\#_t$ could become much bigger through resonance effects.

Contributions involving the extended-technicolour portal arise in all technicolour models that rely on this mechanism (including the scalar versions), as all of them need the four-fermion interaction to generate the standard-model fermion masses. Hence, in models in which the techniquarks carry ordinary colour both contributions appear. In that case they cannot simply be added, as the emission occurs at the same virtualities, and a mutual influence between the emitted technigluons should be considered. 


\subsubsection{More gauge bosons\label{sec:etc}}

We conclude this subsection with a comment on the potential participation also of the lightest extended technigluons, i.e., the ones that provide the masses for the heaviest generation of standard model fermions. They are expected to have a mass of a few TeV. One has to bear in mind that every emission in the gauge-boson ladder has to occur at a (much) larger scale than that. Hence, even in ultrahigh-energy cosmic-ray hadron-hadron collisions the available energy does not suffice for a large number of emissions. In other words, there is not enough energy available to form a dense system of the lightest extended technigluons even at GZK-cutoff energies. Thus, they do not contribute significantly to the hadron-hadron scattering cross section at the GZK cutoff. (This could have been a rather direct contribution, as in models where the techniquarks do not carry ordinary colour the extended technigluon sector unifies directly with quantum chromodynamics.)


\subsection{Cosmic-ray particles beyond the standard model\label{sec:bsm}}

So far, we have only considered effects of the presence of \DEWSB~in modifications of the hadron-hadron cross section. Such extensions of the standard model do, however, not only come with additional gauge interactions, but, in general, also with additional matter. A part of the cosmic-ray particles we observe could be made up from this matter. This deserves an investigation for at least two reasons. As already mentioned at the beginning, so far it has not been possible to find a satisfactory fit to cosmic-ray data by assuming that it is composed of hadrons and nuclei. This might be due to the commonly used straightforward extrapolation of the nuclear cross sections. The cosmic rays might, however, also contain non standard model particles. On the other hand, there is no consensus as to whether a sharp GZK cutoff is observed or not. If there were  trans-GZK events that cannot come from local sources, we must explain, how these high-energy particles make it to us from their far-away sources without being affected by the GZK cutoff. 
If there were none this could also mean that there is no acceleration mechanism that could reach trans-GZK energies. A true GZK cutoff could be distinguished from a limited acceleration mechanism by detecting the particles (and/or their decay products) radiated off the cosmic-ray particles when they are decelerated from energies above to energies below the cutoff by scattering on the CMB photons.

Taken the other way round, if it should turn out that there really are no trans-GZK events {\sl and} if secondaries from the GZK mechanism are found---such that the acceleration to higher energies is possible---this puts limits on the abundance and/or stability of GZK-blind particles.

At the same time incident particles must be sufficiently strongly interacting with the atmosphere to be detected and also in order to be accelerated to high energies in the first place. (Standard model neutrinos, for example, do not interact strongly enough with the atmosphere, which in turn would require huge fluxes.)


\subsubsection{Bound states\label{sec:bs}}

Technicolour theories feature technihadrons. If they are sufficiently long lived to make it from the accelerating source to our atmosphere they can contribute to the cosmic-ray spectrum. The lightest technihadrons are typically the technipions because of their pseudo Nambu-Goldstone nature. In theories with techniquarks that transform under real representations of the technicolour gauge group there are also technibaryons among the technipions. If the technibaryon number is conserved by the extended technicolour sector the lightest technipion is stable and hence sufficiently long lived to make it from the source to the atmosphere. 
For a technihadronic technipion we would at least expect the mass of the Z boson or above, which brings us in the mass range of a heavy nucleus, e.g., iron or beyond.
There can be plenty of other technihadrons, and they all have in common that they have valence techniquarks. There can also be bound states that do not have a direct analogue in the spectrum of quantum chromodynamics. In models where the techniquarks are in the adjoint representation, for example, there can also be bound states of techniquarks with technigluons, which at low energies resemble heavy leptons. 

The sensitivity to a GZK-like cutoff is already altered for electrically charged technicolour bound states: The most potent mechanism for the energy loss of protons is the formation of a baryon resonance and the subsequent radiation of pions. The excitation spectrum in a technicolour bound state is (at least na\"{\i}vely) three orders of magnitude harder than in a bound state of quantum chromodynamics. Hence, such a channel must have a much higher threshold. Furthermore, if the cosmic-ray particle is already the lightest (and sufficiently stable) technicolour bound state, no lighter state can be radiated in the way necessary to cause a standard GZK bound. With this channel at least delayed to higher energies for the charged cosmic-ray particle there remains the much less potent channel for electron-positron production. Moreover, for an electrically neutral technicolour bound state not even this channel is present until the electric dipole moment of possible electrically charged constituents becomes relevant. Independent of the details a delay of the corresponding GZK cutoff to higher energies makes technicolour bound states good candidates for messengers for trans-GZK events. On top of that, due to the enhancement of their cross section at high energies, they can be accelerated efficiently at the source. 

Given that technihadrons will not see the GZK cutoff faced by ordinary hadrons, observing the absence of trans-GZK events can put a limit on the flux of technihadrons, provided at the same time it can be ascertained that the GZK mechanism is really behind the cutoff. (Alternatively, there could simply be no accelerator available that could bring particles to higher energies, which would be a coincidence. The GZK mechanism predicts the production of secondary particles, which must be detected to show that it is at work.) The limit on the flux of technihadrons puts a limit on their stability. Technihadrons can also be dark-matter candidates for which they have to be sufficiently stable. Thus, the above provides another criterion for their viability as dark matter.

 \begin{figure}[t]\begin{center}
 \resizebox{!}{8cm}{\includegraphics{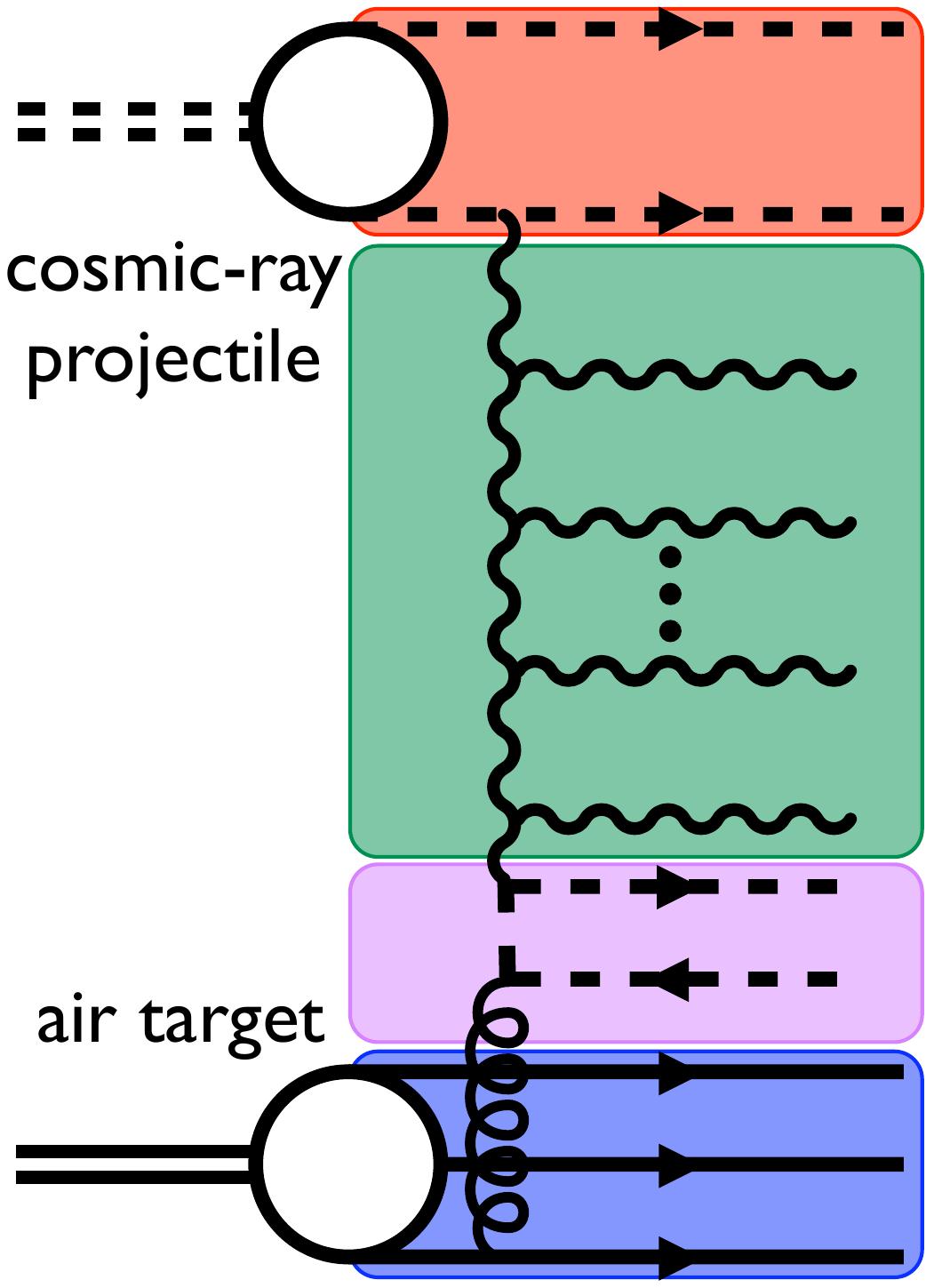}}
 \caption{Technicolour bound-state projectile (top box, red, valence techniquarks shown, dashed lines) scattering with a hadron target (bottom box, blue, valence quarks shown, solid lines). The technigluon ladder (wavy lines, middle box, green) is linked to the valence techniquarks in the projectile and to a pair of techniquarks (third box, lavender) which originate from an ordinary gluon emitted by the valence quarks of the target. (In this sketch, they carry ordinary colour.  Alternatively the extended-technicolour portal contributes. See Fig.~\ref{fig:tc2}.) The ladder is linked to the hadrons through the extended-technicolour sector, i.e., the analogue of the top Yukawa coupling. For techniquarks that carry ordinary colour all combinations of top quark portals (See Fig.\ref{fig:tc1}.) with extended-technicolour portals at the target and projectile sides, respectively, contribute.}
 \label{fig:nonsmproj}
 \end{center}\end{figure}
%
The principal difference when a high-energy technihadron strikes a nucleus in the atmosphere is that the suppression factor for the 
coupling of the projectile to the technigluon ladder is absent, because
the technihadron is already a technidipole in the valence configuration. To the ordinary target hadron, the technigluon ladder connects as before.
(See Fig.~\ref{fig:nonsmproj}.) As a consequence, the total cross section for the liberation of a techniquark from the incident technihadron by scattering on an atmospheric nuclei is given by Eqs.~(\ref{eq:sigmatottc}) and (\ref{eq:sigmatottcrun}), respectively, but without the overall factor of
$\#$, the dependence on $\#$ remaining merely in the logarithm,
\bea
\sigma_\mathrm{tot}^\mathrm{TC^x}
&\approx&
2\pi(\tilde R+ac\bar{\alpha}_\mathrm{TC}y)^2 ,\\
\sigma_\mathrm{tot}^\mathrm{TC^x}
&\approx&
2\pi(\tilde R+a\sqrt{2c\bar{\beta}_\mathrm{TC}y})^2 ,
\eea
where $\tilde R$ is defined in Eq.~(\ref{eq:tilder}) and $\#$ in Eq.~(\ref{eq:hash}).
Therefore, for the benchmark values of the parameters and $N_f=2$ the cross section is increased by a factor of about 23.
That would also increase the expected value for the produced transverse momentum by the same factor.\footnote{
In technicolour theories the longitudinal modes of the weak gauge bosons are technipions, i.e., bound states. They are not stable and therefore are not candidates for projectiles. They can, however, appear in processes with other projectiles. If one of them is emitted from a valence quark in a hadronic projectile it does not face the suppression factor $\#$, but is down by one power of the coupling $\bar{\alpha}_\mathrm{L}\approx1/46$. Then the longitudinal mode can be interpreted as technidipole to which the technigluon ladder connects. 
Hence, the longitudinal modes of the weak gauge bosons represent a third 
portal through which the technigluon ladder can connect to hadronic projectiles and targets.\\ 
This portal is also potent for leptons including the standard model neutrinos. Hence, in the technicolour context at high enough energies, the latter couple much more strongly to a hadronic atmospheric target than in the standard model.
}

At the same time the QCD contribution to the cross section is suppressed, even if we consider
a theory with techniquarks that carry ordinary colour: There a gluon ladder could connect directly to a techniquark in a technicolour bound state. The scale in this bound state is, however, set by the technicolour scale. Hence, the corresponding ordinary colour dipole is extremely small and needs very high momenta to be resolved. Even for the large available rapidity the saturation momentum of quantum chromodynamics does not reach this scale. As a consequence, the hard momentum scale is set by $\Lambda_\mathrm{TC}$. There the coupling of quantum chromodynamics has already evolved to very small values, 
$\bar{\beta}_\mathrm{QCD}/\ln(\Lambda_\mathrm{TC}^2/\Lambda_\mathrm{QCD}^2)$, 
compared to the one of technicolour such that here 
$\log_{10}(Q_\mathrm{QCD}/\Lambda_\mathrm{QCD})\approx1.5$, 
while
$\log_{10}(Q_\mathrm{TC}/\Lambda_\mathrm{TC})\approx3$.
Another contribution that also would proceed through a gluon ladder and which would also be present irrespective of whether the techniquarks carry ordinary colour or not uses the top-quark extendend-technicolour portal on the projectile side. The scale for the coupling of quantum chromodynamics would, however, still be set by the fundamental scale $\Lambda_\mathrm{TC}$ of technicolour and the enhancement factor would still be comparatively small. 

Thus, for a technihadronic projectile the QCD contribution to the cross section is reduced, and this reduction is not compensated by the enhanced technicolour contribution. As a consequence, the total cross section is smaller than for an ordinary hadron. Therefore, the typical height of the first interaction of the cosmic-ray with the atmosphere moves down. At the same time the typical transverse momentum scale is set by the technicolour scale $\Lambda_\mathrm{TC}$ and consequently much larger than for an ordinary hadron as projectile. Hence, technihadronic projectiles can initiate very vehement showers deeper inside the atmosphere, where projectiles, which interact with full unsuppressed QCD strength cannot do so, as their cross section is too big at these energies.\footnote{Thanks are due to Glennys Farrar for pointing this out.}


\subsubsection{Heavy leptons}
 
 \begin{figure}[t]\begin{center}
 \resizebox{!}{7.5cm}{\includegraphics{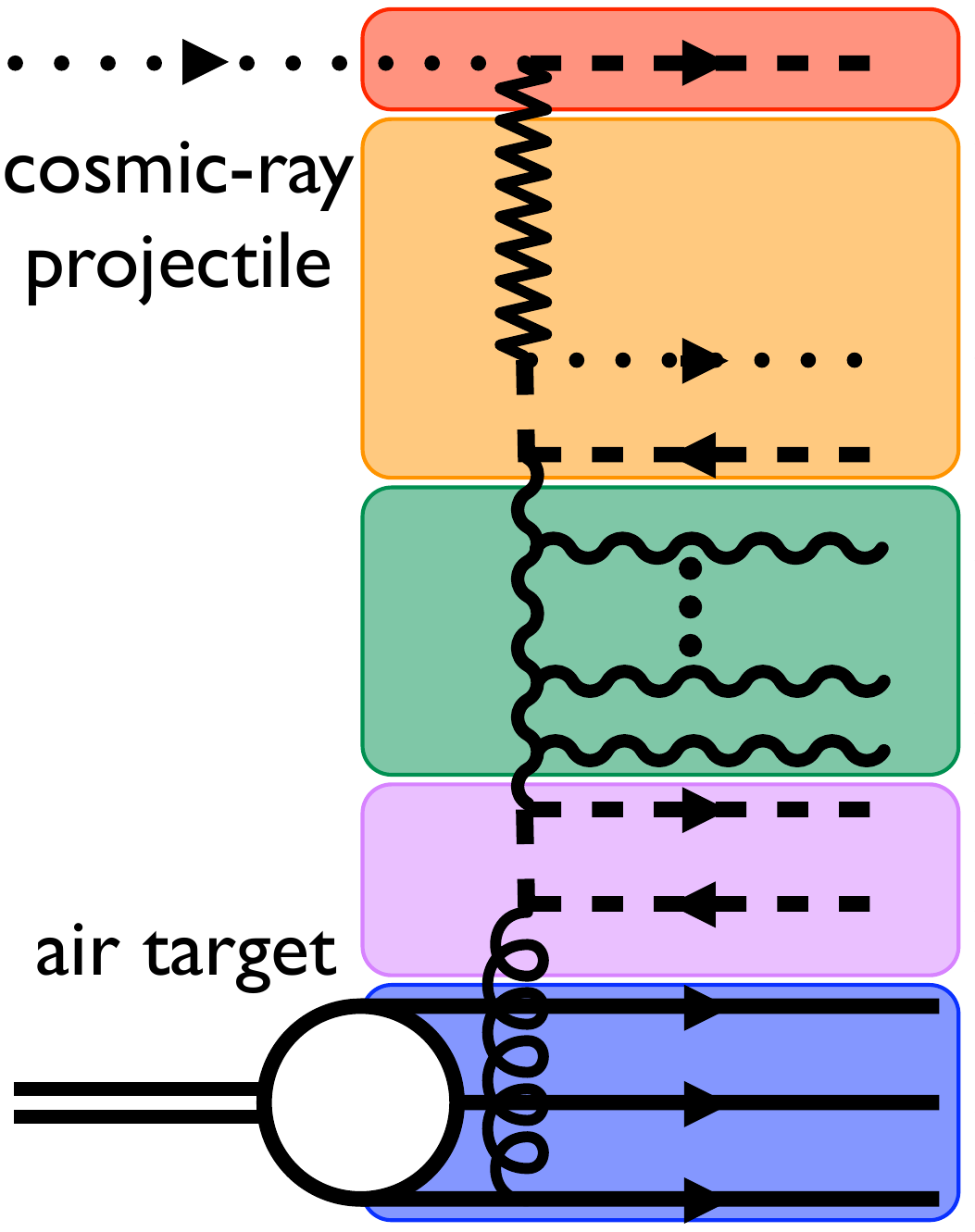}}
 \caption{Dominant contribution from technicolour to the scattering of a heavy lepton projectile (top box, blue, dotted line) on a hadronic target (bottom box, red, valence quarks shown, solid lines). The technigluon ladder (wavy lines, middle box, green) connects to the projectile through the extended technicolour sector, which links the heavy lepton to the techniquarks (second box, orange; the zigzag line represents the extended technigluon). To the target it connects either through the top-quark portal (shown, fourth box, lavender), requiring techniquarks that carry ordinary colour, or through the extended-technicolour sector (as shown in Fig.~\ref{fig:tc2}).}
 \label{fig:nu}
 \end{center}\end{figure}
%
Technicolour models with an odd number of fermion generations charged under $SU(2)_\mathrm{L}$ require a family of additional leptons in order to cancel the Witten anomaly \cite{Witten:1982fp}. Data tells us that these leptons must be heavy (at least about 100GeV \cite{Beringer:1900zz}) to have escaped observation so far. Therefore, like the top quark, they couple strongly through the extended technicolour sector, assuming of course that there is not yet another mass generation mechanism. (See Fig.~\ref{fig:nu}.) Below the mass of the extended technigluon, in analogy to the top-quark portal for an incident ordinary hadron, we get a factor of $Q^4m_{\nu}^2/\Lambda^6_\mathrm{TC}$. At variance with the top-quark portal we do not have the suppression from the emission of the intermediary gluon and the top. 
This would make also heavy neutrinos in the technicolour context to candidates for messengers for trans-GZK events. Among others, one could imagine a scenario where there is a flux of these neutrions (similarly of technihadrons), smaller than the overall effective particle flux below GZK energies and as such hidden. Then if the GZK cutoff is effective for all other species apart from the heavy neutrino (and/or an appropriate technicolour bound state) its small flux would show up in trans-GZK events. Moreover, the enhancement of their cross section at high energies would also allow for an efficient acceleration at the source. (The above is different from the Z-burst scenario \cite{Weiler:1982qy}, where a high energy neutrino annihilates with a relic anti-neutrino in the halo of the earth, sending out a high-energy Z boson that interacts with the atmosphere.)


\subsubsection{Non standard model target}

 \begin{figure}[t]\begin{center}
 \resizebox{!}{6.5cm}{\includegraphics{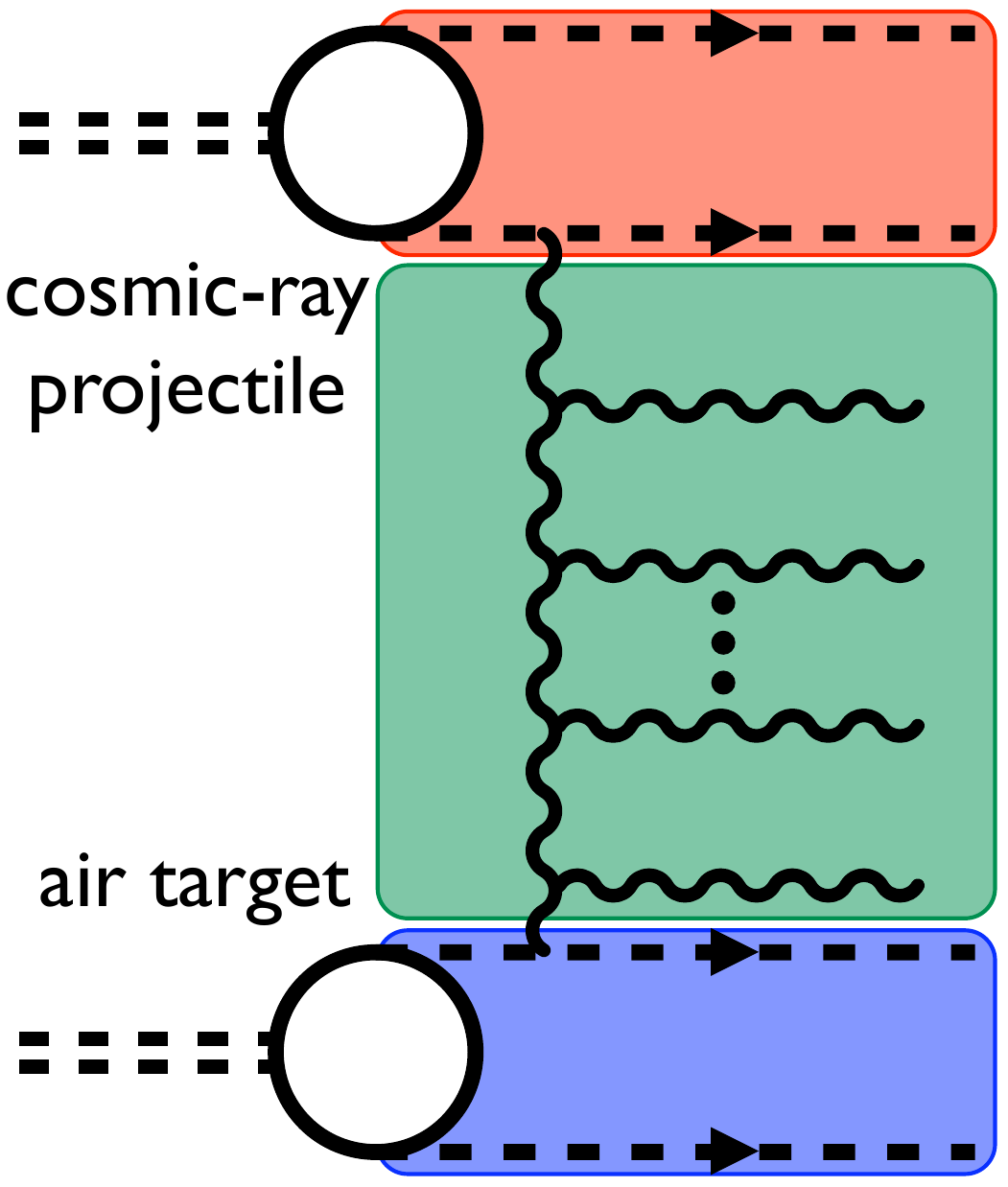}}
 \caption{Scattering of two technicolour bound states [top (red) and bottom (blue) boxes, respectively, valence techniquarks shown, dashed lines]. The technigluon ladder (wavy lines, middle box, green) connects directly to the valence techniquarks.}
 \label{fig:nonsmtar}
 \end{center}\end{figure}
%
It is conceivable that the struck particle is not a standard model particle either, but one arising in an extension.
These targets must interact sufficiently weakly at low energies in order to have eluded direct detection so far, and they must be sufficiently stable to have survived in sufficient numbers. Hence, electrically neutral particles would be likely candidates here. In the context of non--standard-model projectiles we have just mentioned neutral technihadrons, especially technipions, and heavy neutrinos in the technicolour context.
Generally, these particles face constraints like dark-matter candidates. Therefore, let us use the maximal local dark-matter density as an estimate for their abundance, i.e., $\rho_\mathrm{DM}\approx 1 \mathrm{GeV}/\mathrm{cm}^3$. At high energies, when their substructure is resolved, two technihadrons interact like two hadrons. The hadronic interaction length is about $10^2$g/cm$^2$. If it were the same for technihadrons, this would correspond to a physical length of 20Mpc. Thus, this process does not play a role for air showers. It could contribute to the attenuation of this cosmic ray component; after all, the attenuation length for protons due to the GZK effect is about 50Mpc. The halo of the Milky Way is, however, only of the order of 10kpc wide, outside of which the dark-matter density is much lower. Thus, the attenuation would take place on extragalactic scales.

For a technihadronic target (See Fig.~\ref{fig:nonsmtar}.) the technigluon ladder can connect to both the projectile and the target without any suppression. For technicolour theories in which the techniquarks do not carry ordinary colour the contribution from quantum chromodynamics would be even more suppressed, as the ordinary gluons can only connect to the valence techniquarks through the extended-technicolour sector. 
Even for theories in which the techniquarks do carry ordinary colour, the parton distribution functions in the bound states are governed by the technicolour scale and render the coupling of quantum chromodynamics small compared to that of technicolour. For this reason also there the contribution from technicolour dominates the contribution from quantum chromodynamics, and the beginning of the BFKL evolution in the technigluon sector is again not hidden by the contribution from quantum chromodynamics.

The dominant contribution to the cross section, which is induced by the exchange of the technigluon ladder, is given by
\be
d\sigma_\mathrm{tot}^\mathrm{TC^x_x}/d^2b=2(1-e^{-\pi^2Q_\mathrm{TC_x}^2/N_c\Lambda^2_\mathrm{TC}}).
\ee
As parameters for the thickness function $T=T(b)$, which enters the initial conditions for the evolution of $Q_\mathrm{TC_x}$, we choose the $R$=0.5fm for an ordinary proton but rescaled by the ratio $\Lambda_\mathrm{QCD}/\Lambda_\mathrm{TC}$, i.e., $\xx R=R\times(\Lambda_\mathrm{QCD}/\Lambda_\mathrm{TC})$. We do the same for all other corresponding length parameters. (This is consistent with the scale for the tail of the nuclear density being set non-perturbatively by the mass of the lightest hadron, i.e., the pion \cite{Kozlov:2002my}, which in technicolour is raised to the mass of the weak gauge bosons.) For fixed gauge coupling the integration over the impact parameter yields
\be
\sigma_\mathrm{tot}^\mathrm{TC^x_x}\approx2\pi\left(\xx R+\xx ac\bar{\alpha}_\mathrm{TC}y+\xx a\ln\frac{\pi^3\sqrt{2\pi\xx a\xx R}\xx n}{N_\mathrm{TC}\Lambda_\mathrm{TC}^2}\right)^2,
\ee
while for running coupling we find
\be
\sigma_\mathrm{tot}^\mathrm{TC^x_x}\approx2\pi\left(\xx R+\xx a\sqrt{2c\bar{\beta}_\mathrm{TC}y}+\xx a\ln\frac{\pi^3\sqrt{2\pi\xx a\xx R}\xx n}{N_\mathrm{TC}\Lambda_\mathrm{TC}^2}\right)^2 .
\ee
The typical bound state radius makes these cross sections a priori smaller than the hadron-hadron cross sections by a factor of $(\Lambda_\mathrm{QCD}/\Lambda_\mathrm{TC})^2$. This does not impact the question whether the formed system is dense in the same way, as this depends on whether (here at fixed coupling)
$
c\bar{\alpha}_\mathrm{TC}y
>
\ln\frac{N_\mathrm{TC}\Lambda_\mathrm{TC}^2}{\pi^3\sqrt{2\pi\xx a\xx R}\xx n}
$
from which the overall power of the ratio of the scales drops out. There remain additional powers of said scale inside the logarithm that lead to the condition
$
c\bar{\alpha}_\mathrm{TC}y+\ln\frac{\Lambda_\mathrm{TC}^2}{\Lambda_\mathrm{QCD}^2}>\ln\frac{N_\mathrm{TC}\Lambda_\mathrm{TC}^2}{\pi^3\sqrt{2\pi a R} n}
$,
which amounts to a strong enhancement of the density. In fact, it is parametrically the same as for proton-proton scattering through the exchange of ordinary glue with $\sigma_N=\pi/\Lambda_\mathrm{QCD}^2$.
(The analogous condition for a running coupling reads
$
\sqrt{2c\bar{\beta}_\mathrm{TC}y}+\ln\frac{\Lambda_\mathrm{TC}^2}{\Lambda_\mathrm{QCD}^2}>\ln\frac{N_\mathrm{TC}\Lambda_\mathrm{TC}^2}{\pi^3\sqrt{2\pi a R} n}
$.)

Furthermore, these bound states are much heavier than a proton; they naturally have the mass of a Z or a top and can be even heavier. For a given laboratory-frame energy of the incident cosmic-ray particle this fact alone leads to an enhancement of the centre-of-mass energy, as it is approximately the geometric mean of the mass of the (approximately fixed) target and the laboratory-frame energy of the projectile. Hence, for a target mass of 100GeV we would gain one order of magnitude in centre-of-mass energy relative to a proton of about 1GeV. This enhancement of energy would allow a further growth of the cross section.
There arises also a qualitative difference. In Sec.~\ref{sec:etc} we had mentioned the contribution from the lightest extended technigluons, i.e., those responsible for the mass of the heaviest standard-model fermions. While in hadron-hadron scattering even at the GZK cutoff a rapid growth of the contribution should be hampered by the limited available energy, which is needed to emit enough extended technigluons with sufficiently high virtuality, the additional order of magnitude in centre-of-mass energy gained when a target is struck that is a bound state of the technicolour sector counteracts this obstacle. At the scales, where this becomes important, technicolour unifies with the aforesaid extended technigluons. At the corresponding scales, this increases the coupling $\bar\alpha_\mathrm{TC}=\alpha_\mathrm{TC}N_\mathrm{TC}/\pi$ which enters the growth of the gauge boson density through an increase of the number of technicolours $N_\mathrm{TC}$ to the number of colours when it unifies into the lowest-scale extended technicolour sector. As for the expected value for the produced transverse momentum, the multiplication by the scale $\Lambda_\mathrm{TC}$ compensates only for one of the two negative powers from the overall scale.

 \begin{figure}[t]\begin{center}
 \resizebox{!}{8cm}{\includegraphics{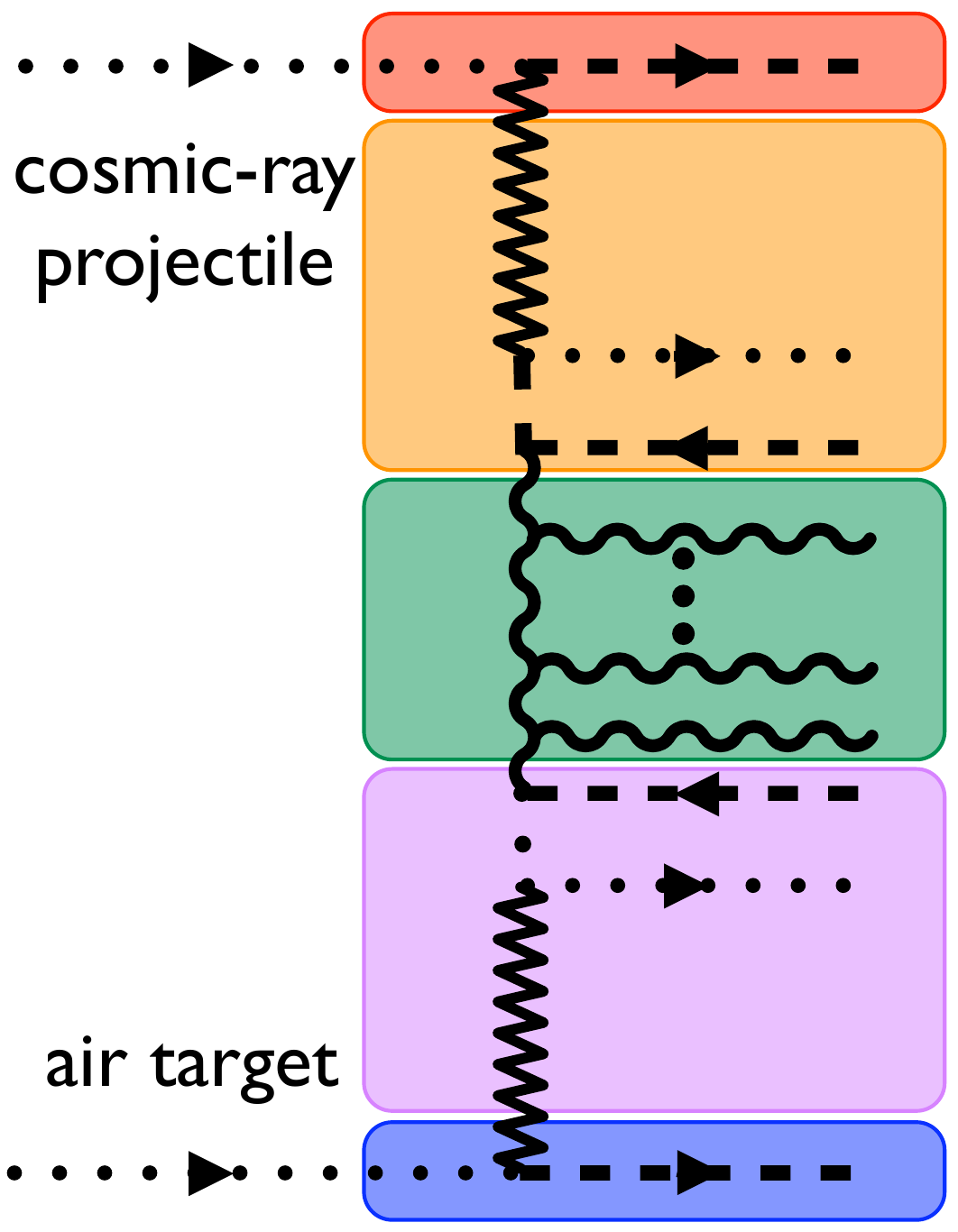}}
 \caption{Dominant contribution to the scattering of two heavy neutrinos [top (red) and bottom (blue) boxes, dotted lines] in the technicolour context. The technigluon ladder (wavy lines, middle box, green) connects to the leptons through the extended-technicolour sector [second (orange) and third (lavender) boxes, zigzag lines represent extended technigluons and dashed lines techniquarks].}
 \label{fig:twonu}
 \end{center}\end{figure}
%
Any combination of technihadrons and heavy neutrinos, respectively, as projectile and target, respectively, can occur. In the technicolour context, the dominant contribution to the scattering of two heavy neutrinos would come from the exchange of a technigluon ladder that connects to the leptons through the extended-technicolour sector, which leads only to a mild suppression due to the analogue of a large Yukawa coupling. (See Fig.~\ref{fig:twonu}.) The density of technidipoles in the neutrinos goes like $\sharp_\nu=m_\nu^2Q^4/\Lambda_\mathrm{TC}^6$ with a typical size of the order $\pi/Q^2$. The decay of their tail is again set non-perturbatively by the inverse mass of the technipions. Hence, as in the previous case we consider $\xx R\approx 0.5 \mathrm{fm}\times\Lambda_\mathrm{QCD}/\Lambda_\mathrm{TC}\approx\xx a$. This leads to
\be
\sigma_\mathrm{tot}^\mathrm{TC^\nu_\nu}
\approx
{\sharp}_\nu2\pi\left(\xx R+\xx ac\bar{\alpha}_\mathrm{TC}y+\xx a\ln\frac{\pi^3\sharp_\nu\sqrt{2\pi\xx a\xx R}\xx n}{N_\mathrm{TC}Q^2}\right)^2 .
\ee
Compared to the previous case, saturation is slightly delayed owing to the elementary nature of the lepton as opposed to the previously regarded technicolour bound state; the scale for the transverse size is not set by the fundamental scale $\Lambda_\mathrm{TC}$ itself, but we have to look at which size ($\sim Q^2$) of dipoles are contained in the wave function of the lepton and $Q^2$ is necessarily above $\Lambda_\mathrm{TC}^2$ if the exchange of a technigluon ladder is considered. The running coupling case is again obtained by the replacement $c\bar{\alpha}_\mathrm{TC}y\rightarrow\sqrt{2\bar\beta_\mathrm{TC}y}$. The top-quark portals do not only lead to technicolour dipoles but also to dipoles of ordinary colour to which an ordinary gluon ladder can connect. The contribution is once more reduced relative to the technicolour contribution because the large momentum scale $Q^2$ leads to a very small coupling in this sector,
\be
\sigma_\mathrm{tot}^\mathrm{QCD^\nu_\nu}
\approx
{\sharp}_\nu2\pi\left(\xx R+\xx ac\frac{\bar{\beta}_\mathrm{QCD}}{\ln\frac{Q^2}{\Lambda_\mathrm{QCD}^2}}y+\xx a\ln\frac{\pi^3\sharp_\nu\sqrt{2\pi\xx a\xx R}\xx n}{N_\mathrm{TC}Q^2}\right)^2 .
\ee

~

\begin{figure}[t]\begin{center}
 \resizebox{!}{8cm}{\includegraphics{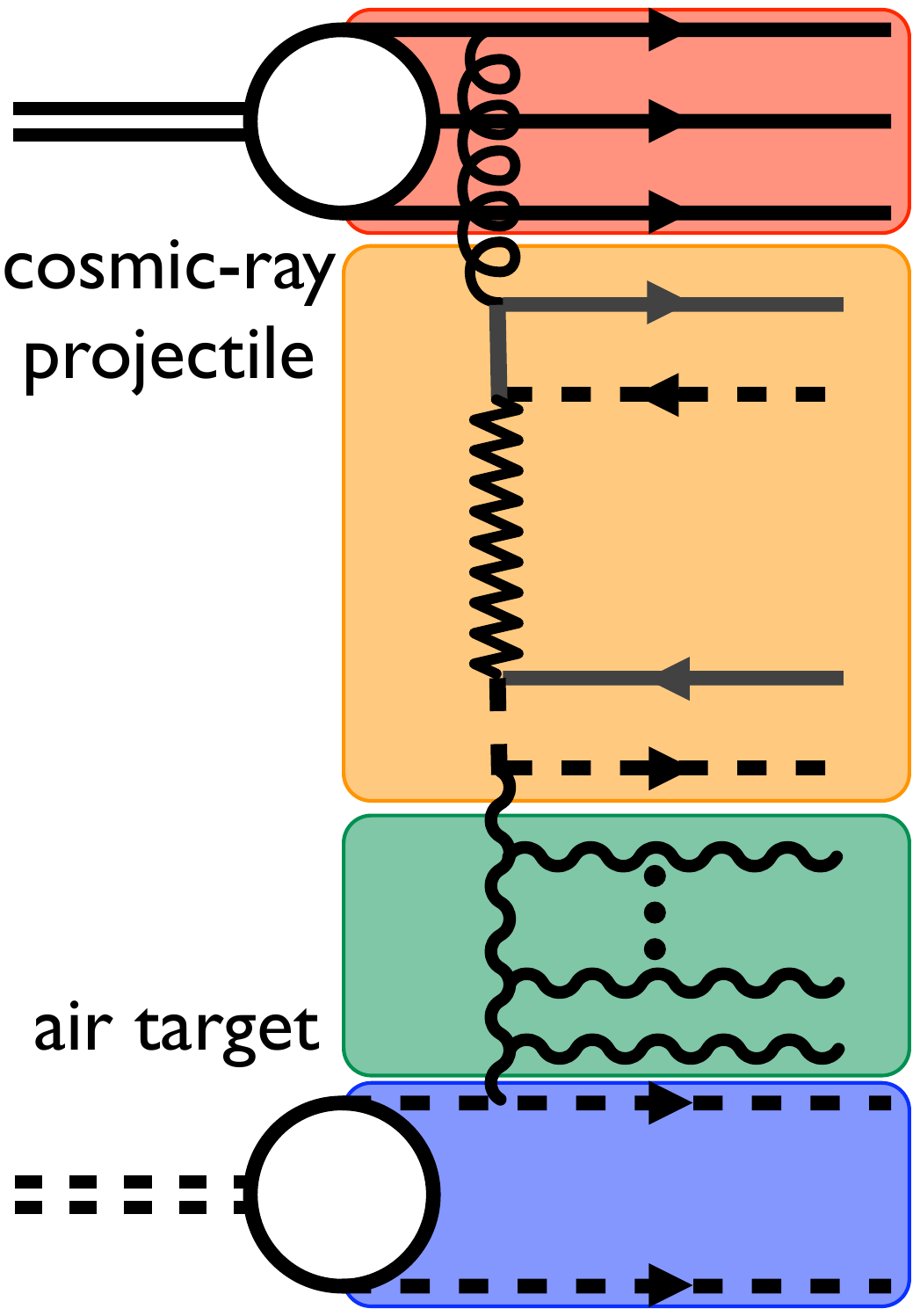}}
 \caption{Scattering of an ordinary hadron [top (red) box, valence quarks shown, solid lines] on a technicolour bound states [bottom (blue) box, valence techniquarks shown, dashed lines]. The technigluon ladder (wavy lines, third box, green) connects directly to the valence techniquarks of the target and to the projectile through the analogue of the top Yukawa coupling (second box, orange; the zigzag lines represent the extended technigluons).}
 \label{fig:smproj}
 \end{center}\end{figure}

The heavy target can also be struck by a hadronic cosmic-ray projectile. (See Fig.~\ref{fig:smproj}.)
%
%
This leads to a suppression of the contribution to the cross section from the exchange of technigluons relative to the previous case, as they must connect to the projectile either through the top-quark portal or, if some of the techniquarks carry ordinary colour, through the emission of a techniquark-antitechniquark pair. The cross section for the latter case at fixed coupling for example reads 
\be
\sigma^\mathrm{TC_x}_\mathrm{tot}\approx\#2\pi\left(\xx R+\xx ac\bar{\alpha}_\mathrm{TC}y+\xx a\ln\frac{\pi^3\sqrt{2\pi\xx a\xx R}\xx n}{N_\mathrm{TC}\Lambda_\mathrm{TC}^2}\right)^2 .
\ee
The contribution from the top-quark (extended technicolour) portal is obtained by replacing $\#$ by $\#_t$ from Eq.~(\ref{eq:hasht}). The running case is obtained by replacing $c\bar{\alpha}_\mathrm{TC}y$ by $\sqrt{2c\bar{\beta}_\mathrm{TC}y}$.

In technicolour theories where the techniquarks do not carry ordinary colour the scattering of a hadronic instead of a non-hadronic projectile on a non-hadronic target leads to an enhancement of the contribution from the exchange of ordinary gluons relative to the case with a non-hadronic projectile since the gluons can connect directly to the projectile, while they connect to the target through the extend-technicolour sector. (The contribution to $\sigma_\mathrm{tot}^\mathrm{TC_x^x}$ that is only present for techniquarks that carry ordinary colour, i.e.~where the ordinary gluons can connect directly to the techniquarks, does not change by more than colour factors if a hadronic target is considered instead of the technicolour bound state.) This yields 
\be
\sigma^\mathrm{QCD_xt}_\mathrm{tot}\approx2\pi
\left(\xx R+\xx ac\frac{\bar{\beta}_\mathrm{QCD}}{\ln\frac{\Lambda_\mathrm{TC}^2}{\Lambda_\mathrm{QCD}^2}}y+\xx a\ln\frac{\sharp_t\pi^3\sqrt{2\pi\xx a\xx R}\xx n}{N_\mathrm{TC}\Lambda_\mathrm{TC}^2}\right)^2 ,
\ee
where the scale for the process is again set by the technicolour scale $\Lambda_\mathrm{TC}$ and $\sharp_t=m_t^2Q^2/\Lambda_\mathrm{TC}^6$.


\subsection{Other realisations of dynamical electroweak symmetry breaking}

As explained in the introduction, there are various other realisations of dynamical electroweak symmetry breaking. Topcolour, for example, contains an additional gauge sector made up of the gauge bosons that generate the required four-fermion interaction. They couple directly to standard-model fermions. In pure topcolour, their mass scale must, however, be extremely high to allow for a light enough top quark. In that case, they could not be excited even in ultra-high--energy cosmic ray events. To the contrary, if topcolour is combined with technicolour or a seesaw mechanism they can have a lower scale. A minimal setup for such a model is based on identifying the SU(3) of quantum chromodynamics with the diagonal subgroup of an SU(3)$^2$, where the first two quark generations are charged under one and the third under the other. 

At virtualities below the corresponding unification scale there is still the contribution from the diagonal subgroup, i.e.~quantum chromodynamics, studied at the beginning of Sect.~\ref{sec:main} and depicted in Fig.~\ref{fig:qcd_bfkl}. At virtualities above the mass $m_\mathrm{tg}$ of the heavy gauge bosons, gauge bosons from both SU(3)-s can participate. The density of sufficiently small dipoles in the hadronic target at the unification scale is approximately given by
\be
x\mathcal{G}(x,m_\mathrm{tg}^2)=\frac{\bar{\beta}_\mathrm{QCD}}{\ln\frac{m_\mathrm{tg}^2}{\Lambda_\mathrm{QCD}^2}}AC_F^\mathrm{QCD}\ln\frac{m_\mathrm{tg}^2}{\Lambda_\mathrm{QCD}^2}\approx A .
\ee
[See also Eq.~(\ref{eq:gluedens}).] The density in a proton projectile is obtained by setting $A=1$. Above the unification scale these dipoles can be seen as being made up to one half of one SU(3) and to one half of the other. Thus, we get
\be
\sigma_\mathrm{tot}^\mathrm{tC1}\approx2\frac{1}{2}2\pi
\left(R
+ac\frac{\bar{\beta}_\mathrm{QCD}}{\ln\frac{Q^2}{\Lambda_\mathrm{QCD}^2}}y
+a\ln\frac{\pi^3An\sqrt{2\pi aR}}{2N_cm_\mathrm{tg}^2}
\right)^2 .
\ee
Here, the evolution of the coupling with the $\beta$ function of quantum chromodynamics is implemented consistently up to the unification scale where it matches the couplings of the two different SU(3)-s. Onward from there the two latter couplings evolve slightly faster than before and different from each other due to the quarks being split to the two sectors, but here we neglect this correction.
Accordingly, a dense system is reached for
$
c\frac{\bar{\beta}_\mathrm{QCD}}{\ln(Q^2/\Lambda_\mathrm{QCD}^2)}y
>
\ln\frac{2N_cm_\mathrm{tg}^2}{\pi^3An\sqrt{2\pi aR}}
\approx 
13.5
$
for our benchmark values together with $m_{tg}=700$GeV. 
For $\ln(Q^2/m_\mathrm{tg}^2)=1$ and $\Lambda_\mathrm{QCD}=250$MeV this necessitates $y\gtrapprox46$, which is predominantly due to the small coupling. Hence, for this particular implementation a sizeable contribution from the additional gauge bosons to the cross section appears unlikely. There remain plenty of variations of the topcolour principle to study, but as long as a unification with the coupling of quantum chromodynamics at scales of this order is required, it is not obvious, how they could lead to a significant part of the hadron-hadron cross section. On a heavier target, e.g.~a top pion, the centre-of-mass energy would be increased, but the issue with the small coupling remains.

~

In composite Higgs models the limit $\xi\rightarrow 1$ is referred to as the technicolour limit. There the compositeness scale is as low as in technicolour. Assuming that the binding of the constituents arises from a gauge interaction, it should be possible to resolve it in ultrahigh-energy cosmic ray events in close analogy to the technicolour case. For smaller values of $\xi$ this becomes more and more unlikely. A more concrete analysis can, however, only be carried out after a (partial) ultraviolet completion in the form of a gauge sector has been specified.

~

For little Higgs models one could also assume that the compositeness is realised by new gauge interactions. A concrete analysis would again require specifying a concrete ultraviolet completion. Even before that, i.e.~at lower energies, there are additional gauge fields that are required to generate the Higgs potential, which is needed to break the electroweak symmetry. The experimentally preferred value for their coupling is at least five times larger than the coupling of the SU(2)$_\mathrm{L}$ \cite{ref:littlehiggsrev}. The lower bound on their mass, however, is 2TeV \cite{ref:littlehiggsrev}, which makes it very energetically expensive to form a dense system in a hadron-hadron cosmic-ray event. A heavier target would increase the possibility of seeing these gauge bosons.

~

 \begin{figure}[t]\begin{center}
 \resizebox{7.5cm}{!}{\includegraphics{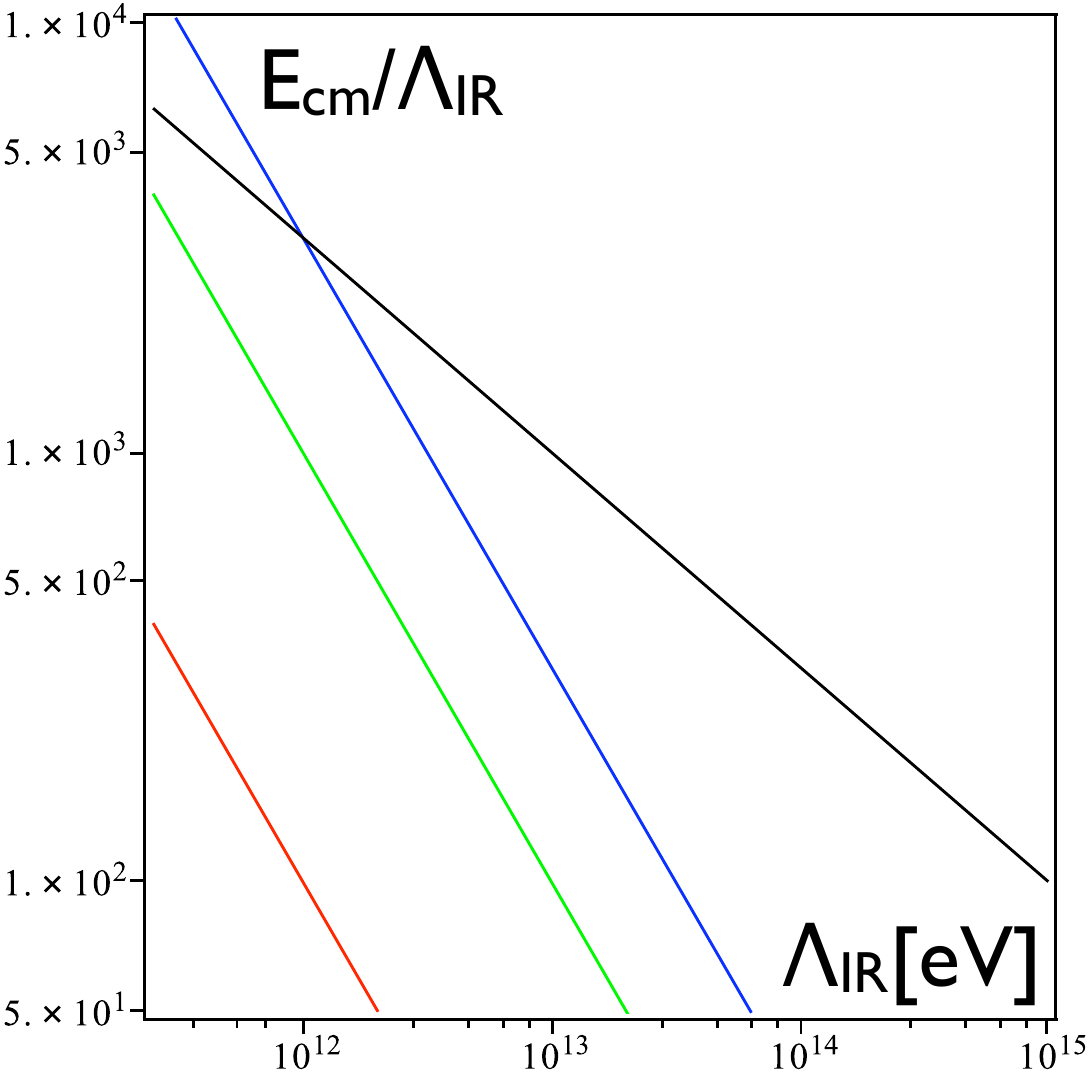}}
 \caption{Ratio of the centre-of-mass energy $E_\mathrm{cm}$ to the infrared scale $\Lambda_\mathrm{IR}$ for different mass scales $M$ for fixed lab-frame energy $E_\mathrm{lab}=10^{19}$eV. The steeper lines from bottom to top are for the scattering by the exchange of a gauge-boson ladder on a proton (red), a technihadron (e.g., a technipion) with a mass of 100 GeV (green), a technihadron with a mass of 1 TeV (blue). Our benchmark values for technigluon ladders are where the lines end on the ordinate. The lines are relevant, for example, for contributions from extended technicolour. The shallower black line is for the scattering by the exchange of a gauge-boson ladder on a target that is a bound state of the same gauge interaction.}
 \label{fig:emissions}
 \end{center}\end{figure}
%
In all of these realisations of dynamical electroweak symmetry breaking there are also candidates for non--standard-model cosmic-ray messengers. These are as diverse as the models they arise in. On top of that, in many cases ultraviolet completions have to be specified in order to be able to carry out the corresponding analyses. Thus, this task must be left for later work. Nevertheless, if the target is much heavier than a proton, an enhanced centre-of-mass energy due to a heavy target may permit the resolution of some of the \DEWSB~mechanisms, which otherwise could not contribute significantly, as not enough emissions are possible for a given centre-of-mass energy and given infrared scale $\Lambda_\mathrm{IR}$. Here, we must distinguish two principal classes of cases. In one class the mass scale for the bound state (and thus the centre-of-mass energy $\sqrt{s}=\sqrt{E_\mathrm{lab}M}$) and the infrared scale for the emissions are independent. For the other class they are coupled. Examples for the first case are: the technigluon ladder ($\Lambda_\mathrm{IR}\leftrightarrow\Lambda_\mathrm{TC}$) for a hadronic target ($M\leftrightarrow\Lambda_\mathrm{QCD}$) or the extended-technigluon ladder ($\Lambda_\mathrm{IR}\leftrightarrow\Lambda_\mathrm{ETC}$) on a technihadronic ($M\leftrightarrow\Lambda_\mathrm{TC}$) or a hadronic ($M\leftrightarrow\Lambda_\mathrm{QCD}$) target. Examples for the second case arise for all gauge sectors that scatter on a target that is a state bound by the corresponding gauge interaction itself. Here the mass and infrared scales change in unison. As a consequence, for fixed lab-frame energy, the centre-of-mass energy is proportional to the square root of the infrared scale and the ratio falls only with $1/\sqrt{\Lambda_\mathrm{IR}}$ instead of $1/{\Lambda_\mathrm{IR}}$ as in the other class of cases. That makes a significant contribution of gauge-boson ladder exchange in such processes possible for much higher scales than one could expect. (See Fig.~\ref{fig:emissions}.)


\section{Summary\label{sec:sum}}

We have studied \DEWSB~in the setting of ultrahigh-energy cosmic rays.
We have considered the possibility that ultrahigh-energy cosmic rays offer the required energy to find direct signs for such a mechanism.

First of all this necessitates evaluating the contribution from within the standard model to ultrahigh-energy cosmic ray events. This contribution is dominated by quantum chromodynamics. 
In QCD, BFKL evolution and subsequent saturation leads to the formation of dense gluonic systems. 
The decisive question is whether quantum chromodynamics already explains the entire cross section deduced from the observation of ultrahigh-energy cosmic rays, e.g., in air showers, or not. There are still considerable uncertainties in the extrapolation of the cross section due to quantum chromodynamics, where we were able to measure it reliably, i.e.~in collider experiments. 

Extensions of the standard model that break the electroweak symmetry dynamically introduce new strongly interacting sectors into our description of the world, which---if they involve additional gauge interactions, and these are the ones, we have concentrated on in this paper---at increasing centre-of-mass energies should share the feature of a rapidly growing cross section with quantum chromodynamics. As a consequence, better knowledge from better data about the cross section of quantum chromodynamics will allow us to better judge on the extensions of the standard model.

In fact, there are two scenarios. Either, based on further studies and measurements, it becomes clear that quantum chromodynamics does explain the entire cosmic-ray--air cross section to a given accuracy in which case we can use this information to rule out extensions that would have further modified said cross section to this accuracy, or it turns out that the cross section cannot be explained by quantum chromodynamics alone, which implies that extra contributions are needed, which could be supplied by additional gauge interactions like the ones present in \DEWSB.

It should also be emphasised that this would be the point where the {\sl ultraviolet} sector of \DEWSB~can be probed and where it can be told apart from an effective description in terms of composite fields in four dimensional or extra-dimensional set-ups.

In our investigation, we have distinguished different systems. We have started by studying the contribution of extensions to the case where the cosmic-ray projectile as well as the target in the atmosphere are ordinary hadrons. There the contribution from the extension is suppressed alone by the fact that, unlike quantum chromodynamics, it cannot connect directly to the constituents of the hadrons. 
Nevertheless, at energies around the ankle of the cosmic-ray spectrum a low-scale extension like technicolour can still form a dense system of gauge bosons (here technigluons).
Below GZK-cutoff energies
the contribution to the hadron-hadron cross section from technicolour will typically be a per mill of the contribution from quantum chromodynamics.  
Then again, technicolour interactions take place at a factor 1000 higher momentum scales,
which could lead to rare but vehement events. If one, however, studies QCD cross sections at these high energies, they are of the same order as the technicolour ones. Therefore, no smoking gun is likely to show up by looking at total cross sections in hadron-hadron interactions. Such a non-qualitative change of the cross section requires a lot of statistics to disentangle, which is not (yet) available at these energies. This is especially true in view of the fact that with present data such changes in the cross section can still be traded against changes in the assumed composition. The uncertainties in composition are the largest source of uncertainties in the analysis of air showers to date \cite{Kampert:2014eja}.
The progression of a shower induced by a technicolour interaction should be different, though, as the dense system is formed by technigluons and not by gluons. The latter feed into hadrons via ordinary quarks. The technigluons cannot decay that directly into standard-model matter and can even produce (meta)stable non-standard matter. The details hereof are more sensitive to the details of the new strongly interacting sector than the aspects studied here, and a corresponding investigation must be left for later work.

Sectors that break the electroweak symmetry dynamically also come with other candidates for the projectiles. These can be bound states of the new strongly interacting sector but also new heavy leptons needed to cancel anomalies. First of all, they do not see the GZK cutoff predicted for ordinary hadrons. (On photons from the cosmic microwave background nuclei photo-disintegrate, while protons are exited and radiate off pions.) The bound states of a new sector are much more deeply bound. Hence, the scale of the cutoff is correspondingly higher. (A similar statement holds for neutrinos.)
What the cross section is concerned, their constituents can couple directly to the technigluons and the corresponding suppression of the cross section drops out. At the same time, quantum chromodynamics receives a two-fold penalty; apart from models where the techniquarks carry also ordinary colour it cannot couple directly to the constituents anymore, and, moreover, the momentum scale set by the bound-state wave function is much larger than the fundamental scale of quantum chromodynamics, which makes its gauge coupling very small. (If one could isolate such events, even the onset of the HERA-like growth of the cross section mediated by the technigluon ladder could be visible.)\\ 
Moreover, the aforementioned reduction of the QCD cross section is not compensated by the increase of the technicolour cross section. Thus, the overall cross section is reduced. Hence, such a highly energetic projectile can initiate the air shower deeper in the atmosphere. To the contrary, a projectile interacting with the full QCD cross section can only penetrate that deeply at smaller centre-of-mass energies, where the QCD cross section is smaller.\\
Furthermore, since the new bound states do not see the ordinary GZK-cutoff, there should be trans-GZK events if these new bound states are sufficiently stable and there is no reason---coincidental (no accelerator in the vicinity) or fundamental (no possible acceleration mechanism)---why they cannot be accelerated to higher energies. (Whether the GZK mechanism is at work for ordinary hadrons, i.e., whether there are actually hadrons decelerated by it from super- to sub-GZK energies, could be checked by trying to observe the secondary particles emitted in the process.)
\\
The heavy leptons couple without big suppression to the technigluon sector through the analogue of a large Yukawa coupling. The direct coupling of all these non-hadronic projectiles to the strongly interacting sector also allows for their efficient acceleration at the source. Light neutrinos offer a different glance at \DEWSB~extensions of the standard model. In the absence of (the analogue) of a large Yukawa coupling they couple democratically to quarks and techniquarks through the exchange of weak gauge bosons. Despite the suppression by the weak coupling this offers a relatively clean look at standard-model and non--standard-model components of the cross section \cite{CooperSarkar:2011gf} also at neutrino observatories such as IceCube \cite{ref:icecube}.

In technicolour the connection between the gauge sector and electroweak symmetry breaking is most direct and consequently the involved energy scales are the smallest possible. In all other realisations of \DEWSB~the connection is less direct and the energy scale for probing the additional gauge sector is delayed. That makes finding an increase of the corresponding contribution to the cross section energetically more and more difficult. Concretely, in topcolour the gauge interactions generate four-fermion interactions, which later break the electroweak symmetry. This is very similar to the extended technicolour sector; the part that gives the masses to the heaviest standard-model fermion generation comes in at the lowest scales. In composite and little Higgs theories a priori no gauge interaction is specified, but under the assumption that the Goldstone sector of these theories is generated in this way, it is detached from the electroweak scale by virtue of the Goldstone theorem. In the so-called technicolour limit of composite Higgs theories where the pion decay constant and the electroweak scale coincide a similar cosmic-ray phenomenology could be expected. A concrete analysis necessitates, however, specifying an ultraviolet completion for these models. This also holds for little Higgs theories only that there are additional gauge fields above at least 2TeV that generate the Higgs potential.

If the target is not a hadron but a bound state of the extension or a heavy lepton, then, to begin with, for a given lab-frame energy the centre-of-mass energy grows like the square root of the rest mass of the target. For the increase from the proton to the Z mass this already provides an additional order of magnitude. Thus, the prospect on discovering new gauge interactions by scattering on bound states they hold together is improved, as for these targets reaching the energy frontier is postponed. (See Fig.~\ref{fig:emissions}.)

There are several paths that could be taken to further elaborate on the above results.
It would be very beneficial to obtain data on the growth of the cross section of quantum chromodynamics with the energy closer to the energies of the ultrahigh-energy cosmic rays but in a clean environment, which can be achieved in proton-nucleus collisions at the upgraded Large Hadron Collider. 
This is needed as better starting point for extrapolating the part of the cross section coming from quantum chromodynamics and as a milestone for understanding the growth of the cross section in an additional gauge sector. Moreover, the above computations can, e.g., be taken to higher orders and/or beyond the mean-field approximation in the initial conditions as well as corrected for quantum fluctuations in the evolution of the parton distribution functions. That will also serve to get a better understanding of how to translate the quantum chromodynamics measurements to new gauge sectors.
Moreover, the findings can then be implemented in a full shower simulation.
Furthermore, here, we mostly compare total cross sections. 
More pieces of information from ultrahigh-energy cosmic-ray observations could be compared to our predictions, like, e.g., the secondary composition of the shower or the shower direction. An investigation of event-by-event fluctuations was already mentioned above, but requires a large number of recorded cosmic-ray events. Combining these investigations with information obtained from considering the acceleration mechanism at the source and/or the transport through space to the earth could also give additional insight \cite{Hooper:2008pm}.


\section*{Acknowledgments}

The author acknowledges gratefully discussions with
Roberto Contino,
Adrian Dumitru,
Glennys Farrar,
Stefan Hofmann,
Martin Holthausen,
Michael Kopp,
Manfred Lindner,
Stefan Pokorski,
Joachim Reinhardt,
Subir Sarkar, 
Karolina Socha,
and Mark Strikman.
This work was supported in part by the
the ExtreMe Matter Institute (EMMI),
the Helmholtz International Centre (HIC) for FAIR within the framework of the LOEWE program launched by the State of Hesse,
and 
by the Humboldt Foundation.


\end{document}